\documentclass[lettersize,journal]{IEEEtran}
\usepackage{array}
\usepackage{booktabs}
\usepackage{tabularx}
\usepackage{multirow}
\usepackage{amsmath,amsfonts}
\usepackage{mathtools}
\usepackage{amsmath}
\usepackage{indentfirst}
\usepackage{titlesec}
\usepackage{array}
\usepackage{textcomp}
\usepackage{stfloats}
\usepackage{url}
\usepackage{verbatim}
\usepackage{graphicx}
\usepackage{subcaption}
\usepackage{cite}
\usepackage{algorithm}
\usepackage[noend]{algpseudocode}
\begin{document}

\title{Evolutionary City: Towards a Flexible, Agile and Symbiotic System}

\author{Xi Chen$^{1,2,\dagger}$,
        Wei Hu$^{1,\dagger}$,
        Jingru Yu$^{1}$,
        Ding Wang$^{1}$,
        Shengyue Yao$^{1,*}$,
        Yilun Lin$^{1,*}$,
        and Fei-Yue Wang$^{3,*}$

%
%

\thanks{This work is supported by the Shanghai Artificial Intelligence Laboratory, grant number 2022ZD0160104.}
\thanks{$\dagger$ Equal contribution.}
\thanks{* Corresponding author: Shengyue Yao (yaoshengyue@pjlab.org.cn), YilunLin(linyilun@pjlab.org.cn), Fei-Yue Wang (feiyue.wang@ia.ac.cn)}
\thanks{$^{1}$Xi Chen (chenxi2331@sucdri.com), Wei hu (huwei@pjlab.org.cn), Jingru Yu (yujingru@pjlab.org.cn), Ding Wang (wangding@pjlab.org.cn), Shengyue Yao, Yilun Lin are with Urban Computing Lab, Shanghai Artificial Intelligence Laboratory, Shanghai, China.}%
\thanks{$^{2}$Xi Chen is with Environment \& Traffic Design and Research Institute, Shanghai Urban Construction Design \& Research Institute (Group) Co., Ltd., Shanghai, China.}%
\thanks{$^{3}$Fei-Yue Wang is with the Institute of Automation, Chinese Academy of Sciences, Beijing, China, and the Macau Institute of Systems Engineering, Macau University of Science and Technology, Macau, China.}

}

\markboth{IEEE Transactions on Intelligent Vehicle, 2023}%
{Shell \MakeLowercase{\textit{et al.}}: Bare Demo of IEEEtran.cls for IEEE Journals}


\maketitle

\begin{abstract}

Urban growth sometimes leads to rigid infrastructure that struggles to adapt to changing demand. This paper introduces a novel approach, aiming to enable cities to evolve and respond more effectively to such dynamic demand. It identifies the limitations arising from the complexity and inflexibility of existing urban systems. A framework is presented for enhancing the city's adaptability perception through advanced sensing technologies, conducting parallel simulation via graph-based techniques, and facilitating autonomous decision-making across domains through decentralized and autonomous organization and operation. Notably, a symbiotic mechanism is employed to implement these technologies practically, thereby making urban management more agile and responsive. In the case study, we explore how this approach can optimize traffic flow by adjusting lane allocations. This case not only enhances traffic efficiency but also reduces emissions. The proposed evolutionary city offers a new perspective on sustainable urban development, highliting the importance of integrated intelligence within urban systems.

\end{abstract}


\IEEEpeerreviewmaketitle

\section{Introduction}


Cities are complex entities that promote a mutual relationship among people, nature, infrastructure, and institutions\cite{lopes2013public}. This relationship represents a delicate balance between the urban residents' demands and the the city resource supplies. Over the past centries, cities evolve along with the disruption and restoration of this balence. From the agriculture-based City 1.0 to the information-driven 4.0, cities have refomed not only physically but also in their cultural, religious, and emotional dimensions\cite{lin2023city, ferrer2017barcelona}.Recent technology advancements in Artificial Intelligents ushers in the era of City 5.0. With dissolved national boundaries, extended regional influences and intensified connections among their residents, cities have evovled into complex systems \cite{button2015global, batty2008size,allen2012cities,portugali2012complexity}. As a consequence, the imbalance between demands and supplies is exacerbated. \cite{bettencourt2013origins, barabasi2013network}.

Two striking phenomena in the era of City 5.0 attract particular attentions. Initially, the occurance of unexpected events grows significantly, which leads to a great potential of system-wide emergencies \cite{ahern2011fail}. This phenomenon is resulted from the emerging demands by AI-driven equipments, which are lack of interpretability \cite{godschalk2003urban, burrell2016machine, yigitcanlar2020contributions}. Furthermore, the supply-demand dynamics between cities and their residents have become increasingly volatile. \cite{batty2013new}. One primary reason for this phenomenon is the sluggish adaption of the city resource supply, which is resulted from the substantial time and capital cost in constructing infrastructures \cite{trubka2010costs}.  

\begin{figure}[ht]
    \centering
    \begin{subfigure}[b]{0.8\linewidth}
        \centering
        \includegraphics[width=0.8\linewidth]{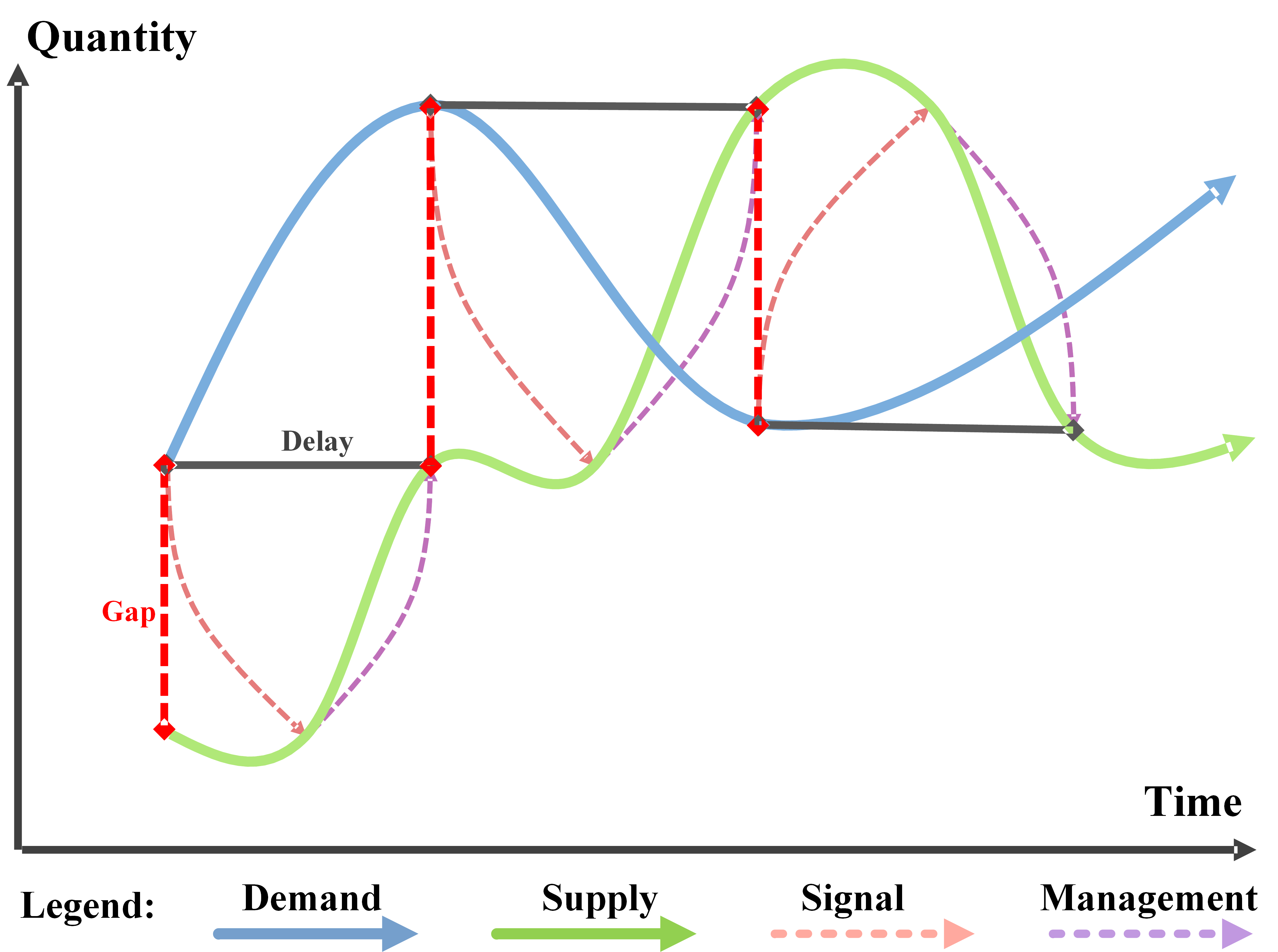}
        \caption{Insufficient flexibility and adaptability in supply adjustments arise from the rigidity within urban systems, ultimately resulting in a significant mismatch between supply and demand.}
        \label{f1}
    \end{subfigure}
    \begin{subfigure}[b]{0.8\linewidth}
        \centering
        \includegraphics[width=0.8\linewidth]{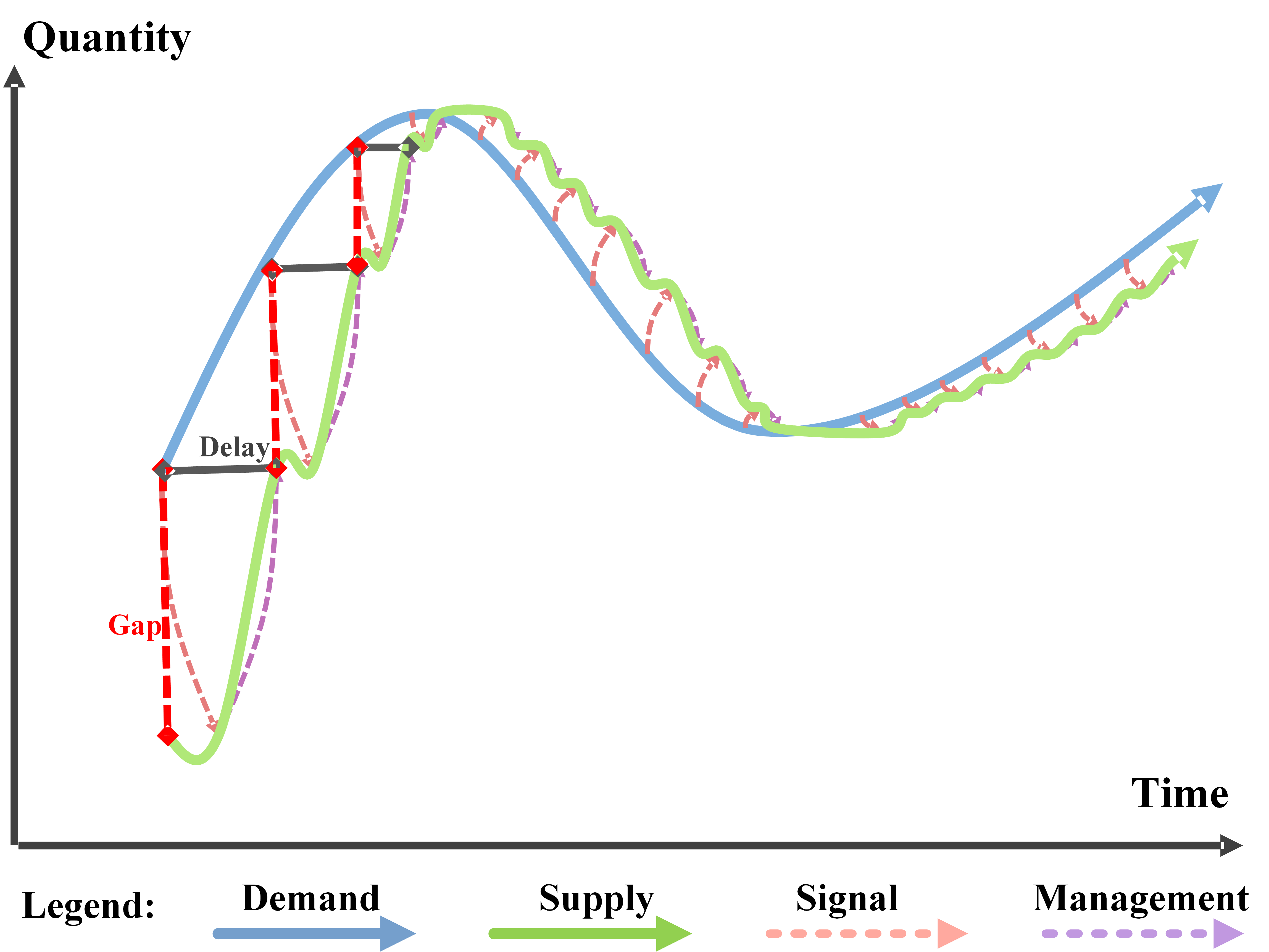}
        \caption{Enhancing the flexibility and adaptability of supply through high-frequency dynamic management, ultimately bridging the gap between supply and demand.}
        \label{f2}
    \end{subfigure}    
    \caption{The effect of supply and demand adjustment with a flexible and adaptable management.}
    \label{f}
\end{figure}

These phenomena proposed an compelling requirement of flexibile and agile urban management approaches. To achieve this requirement, three challenges are ignited comparing Fig. \ref{f1} with Fig. \ref{f2}, which are:

\begin{itemize}
\item Reducing the latency between the perception of environments and prompting demands. 
\item Increasing the estimation/prediction accuracy of supply-demand dynamics. 
\item Improving the decision-making effectiveness and efficiency of AI-driven equipments in the complex urban system. 
\end{itemize}

However, these challenges are intricate to be tackled by conventional urban management approaches. Considering the forthcoming human-machine coexistance ecosystem in the era of City 5.0, these challenges must be tackled in a mutual trust and alignment manner, which provokes further difficulties.



To this end, this paper introduces an evolutionary city framework. The proposed framework utilizes a human-machine symbiotic intelligence mechanism, which can capitalize on the unique strengths of both experts and AIs. By integrating the concepts of adaptability perception, parallel simulation, and autonomous decision-making, urban management can better synchronize supply with demand, as depicted in Fig. \ref{f2}. Specifically, to tackle the issue of delayed demand perception, we utilize fused multi-modal sensor data in conjunction with efficient data cycling. This enhanced perception technique for urban environment and demand serves as an adaptive mechanism contributing to the evolution of urban systems\cite{bettencourt2014uses}. To ensure accurate estimation/predictions of future urban demand, we utilize data-driven AI simulation techniques based on graph neural networks, offering realistic city development simulations and predictions. This dynamic simulation further refines the mutation combination process in the evolution of urban systems.\cite{batty2013new}. Addressing the effectiveness and efficiency challenges of integrated human-machine decision-making, we capitalize on both expert insights and AI capabilities in a decentralized and autonomous organization fashion. This facilitates the rapid selection of urban evolutionary trajectories \cite{neirotti2014current, yao2023towards} and minimizes delays in human-machine collaboration arising from trust discrepancies and misalignments \cite{kasznar2021multiple,gann2011physical}. Under the proposed framework, the city can effectively utilize its existing infrastructure to meet the continuously evolving demand\cite{batty2012smart,neirotti2014current}, rather than expending vast resources on renovating infrastructure.

Taking a specific instance, the implementation of variable lanes in Shanghai Lingkong SOHO exemplifies the pragmatic relevance of the evolutionary city framework. Rather than adhering to the static utilization of the space, the variable lanes methodology intelligently and flexibly modifies the number of lanes in each driving direction. Combining with other intelligent traffic management approaches (e.g., signal control, variable speed limit, etc.), the transportation system can be empowered towards flexible, agile and symbiotic intelligence under the framework of the evolutionary city. 

The primary contributions of this study are as follows:


\begin{itemize}
\item The mechanism of human-machine symbiosis is proposed to be applied to the framework of the evolutionary city, aiming to enhance the mutual trust and information alignment between humans and machines.
\item Adaptability perception, parallel simulation, and autonomous decision-making are incorporated into the evolutionary city framework to address the predominant challenges in urban management.
\item Through the application of the evolutionary city framework to lane management, simulation experiments and practical implementation have demonstrated a reduction in traffic delays by 15\%-20\%.
\end{itemize}

The rest of this paper is organized as follows. Section \ref{relatedwork} investigates the related work. Section \ref{method} elaborates the proposed evolutionary city framework and discusses its pillar technological components. Section \ref{casestudy} elaborates the variable lane management approach and its application, by which the feasibility and efficiency of the proposed framework is demonstrated. Finally, this research is concluded in section \ref{conclusion}.

\section{Related work} \label{relatedwork}

Existing studies regarding enhancing the agility and flexibility in city management are reviewed in this section. Specifically, works on human-machine interaction, perception, simulation, and decision-making in the urban system is discussed, respectively.

\subsection{Human-machine interaction in urban system}

To enhance urban management and planning, it is essential to integrate human-machine interaction into processes such as perception, simulation, and decision-making.
The significance of human-machine interaction in smart city surveillance systems has been highlighted\cite{kashef2021smart-human2}, emphasizing the need for effective interaction between humans and technology to ensure the ethical and responsible use of surveillance technologies. Another study explores city managers' perceptions of artificial intelligence in local governments, highlighting the importance of considering human factors in AI implementation in urban contexts\cite{yigitcanlar2023artificial-human1}. Furthermore, addressing human factors and promoting effective human interaction with decision-making in smart cities is crucial, as demonstrated in a study on human interaction in smart city cases \cite{trende2022case-human}.

Building upon human-machine interaction, the concept of symbiotic intelligence is proposed. Symbiotic intelligence, as defined in previous work\cite{jacucci2014symbiotic}, combines computation, sensing technology, and interaction design to enable deep perception, awareness, and understanding between humans and machines.
This concept further addresses the challenge of human-technology symbiosis in the context of smart cities\cite{mckenna2020beyond}. 
By  incorporating symbiotic intelligence and considering human factors, 
it becomes possible to effectively perceive the dynamic demand, simulate urban evolution, and make informed decisions that align with the values and preferences of their communities in urban management and planning.

\subsection{Urban perception}

Perception technologies allow smart services to monitor the city's adaptability and react to dynamic demands. Sensors are becoming increasingly prevalent and abundant, due to the reduced production cost of high-quality sensors and the widespread embedded nature of sensors within smart devices. This allows for real-time identification of demand and supply distribution within urban systems through sensor networks and data analytics technology.

An illustrative example of perception in supply distribution is the utilization of sensor networks for real-time monitoring of urban infrastructure and facilities.
Smart city programs provide a range of technologies that can be applied to supplier perception associated with aging infrastructure and increasing demands\cite{berglund2020smart}. 
Applying advanced sensors to monitor public infrastructures, such as bridges, roads, and buildings, allows for more efficient use of resources based on the collected data\cite{hancke2012role-sns}.


To enhance supply-demand alignment, a combination of dedicated and non-dedicated sensors is utilized, leveraging sensing technologies and data generation to optimize infrastructure layout\cite{jiang2020federated, habibzadeh2017smart-sns}.  Data collected by these sensors serve as valuable inputs for training advanced Artificial Intelligence (AI) models, benefiting various smart services that cater to societal demands. However, the reliance on legacy data acquisition models and centralized machine learning approaches raises concerns regarding security and privacy. Consequently, there is reduced participation in large-scale sensing and data provision for smart city services. Addressing these challenges, federated Learning emerges as a promising solution, ensuring privacy and security in the data collection process\cite{jiang2020federated}.

\subsection{Urban system simulation}

Transforming the city into a smarter entity is a formidable endeavor owing to the intricate nature of city environments. The city is not an automated system that can be easily understood and predicted, but rather a living system that evolves every day through variations and developments of its physical constructs, economic and political activities, social and cultural settings, and ecological systems. The modeling and simulation technology of the urban system is anticipated to serve as a reliable means of accurately mirroring and influencing the city's functions and processes, thus amplifying its realization, operability, and management\cite{deng2021systematic,shahat2021city-sim,qin2023autosim,li2017parallel,wang2010parallel,chen2023acp,li2023parallel}.

The study of urban system modeling and simulation has advanced greatly over the last few decades, turning into a vital resource for researchers and planners of transportation. Genetic algorithms offer a powerful optimization approach to simulate city evaluation that can be applied to various aspects of urban planning and transportation management, including transit network design \cite{bourbonnais2021transit-evolution}, traffic signal optimization\cite{mao2021boosted-evolution}, and congestion mitigation\cite{sanchez2009traffic-evolution}.

The introduction of AI-driven cars onto our roadways necessitates an altogether new level of modeling complexity, whereas conventional models tend to focus on static elements~\cite{hu_review_2022,watta_vehicle_2021,zhou_car-following_2023,zhang_hivegpt_2023}. 
Early simulations were mathematical and only considered static elements like population density and land use patterns. These simplistic models, however, were replaced by complex simulations integrating cutting-edge algorithms capable of modeling individual vehicle behaviors, traffic flow dynamics, and network optimization as computer technology advanced and the concept of intelligent vehicles became more tangible~\cite{barcelo_fundamentals_2010,lin_mobility_2023}. As intelligent vehicles began to surface, the need to incorporate real-time data and connected vehicle technology became undeniable.
As a result, simulations started to accurately reflect the realities of roadways filled with both human- and AI-driven vehicles~\cite{li_simulation_2023}. Various simulation models such as SUMO \cite{lopez_microscopic_2018}, MATSim \cite{andreas_introducing_2016}, AimSun \cite{aimsun_aimsun_2022}, VISSIM \cite{ptv_vissim_2012}, and others have been developed to simulate traffic systems in the city with diverse scales.



\subsection{Intelligent decision-making for urban system}
Decision-making plays a crucial role in urban system, especially for resource distribution and infrastructure utilization. Traditional decision-making methods involve resource allocation in uniform regions, which might not always be efficient for hard-to-predict situations. 

To address this issue, previous research has put forth an intelligent decision-making framework that evaluates and prioritizes city's resource allocation to facilitate sustainable recovery during the  COVID-19 pandemic \cite{longsheng2022smarter-decision}. Scholars have also explored integrated decision-making approaches in various domains such as energy\cite{shahsavar2022bio-decision}, infrastructure\cite{mei2018integrated-decision}, and logistics\cite{janjevic2019integrated,li2023logistics}, specifically focusing on resource allocation.


The implementation of reversible lane technique based on adaptive traffic management concept serve as a practical case example of how intelligent decision-making strategies can achieve flexible resource allocation and optimization\cite{wang2008toward,oughton2018infrastructure}.
Specifically, in scenarios where inbound and outbound traffic on a facility is unbalanced throughout the day, a lane management strategy called tidal flow or reversible lane control can be beneficial\cite{ampountolas2019motorway-lane}. This strategy offers several benefits compared to traditional network design methods, including improved network performance, optimal resource utilization, ease of operation, quick returns, cost-effectiveness, and no additional land requirements \cite{zhao2019saturation-lane}.

\section{Framework} \label{method}

\begin{figure}[ht]
\centering
\includegraphics[width=1\linewidth]{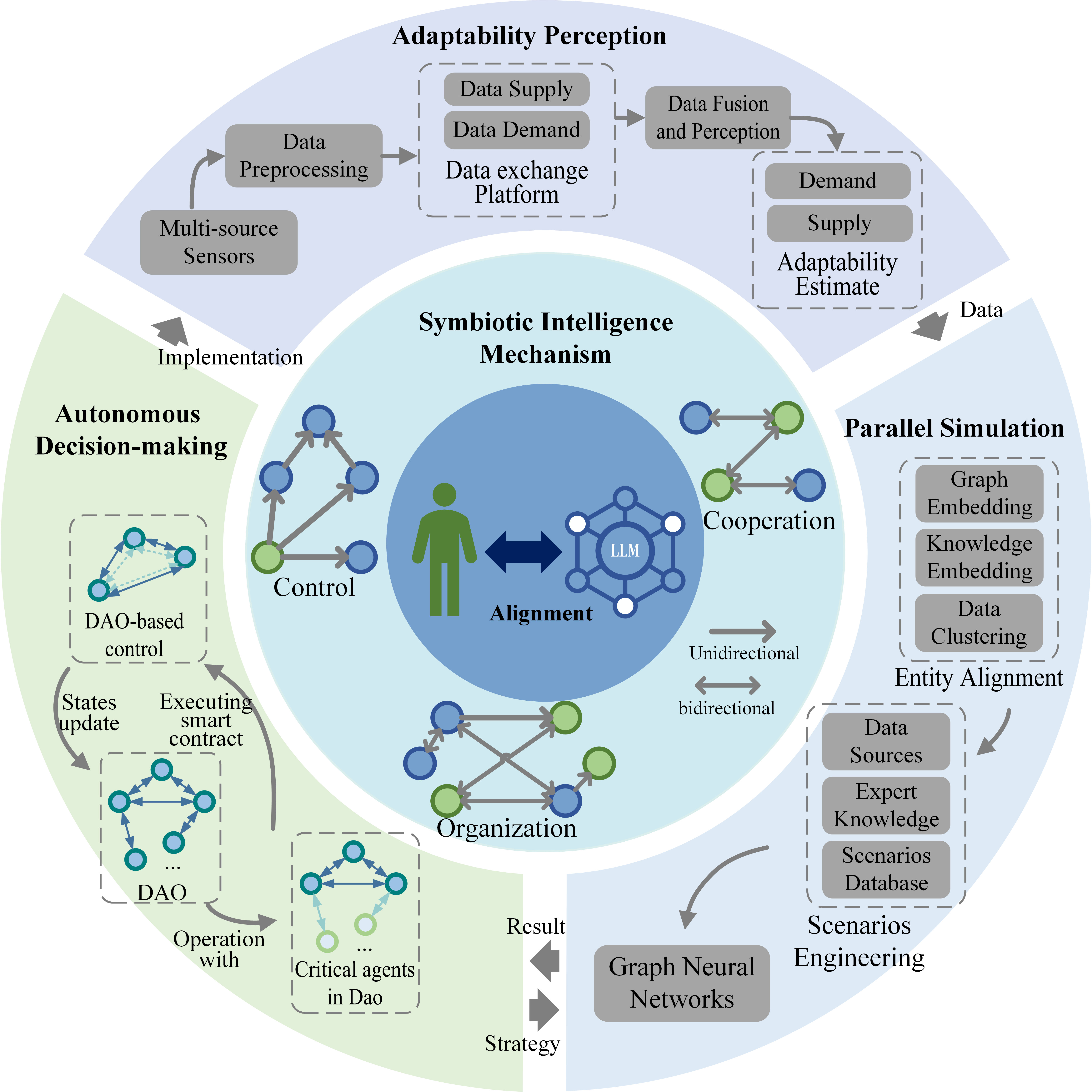}
\caption{The evolutionary city framework. The framework has four key components, adaptability perception, parallel simulation, autonomous decision-making, and symbiotic intelligence mechanism.}
\label{framework}
\end{figure}


The evolutionary city framework is introduced in this section, as illustrated in Fig. \ref{framework}.
Centered on human-machine symbiotic intelligence, the evolutionary city framework leverages human-machine synergy across three pivotal domains: adaptability perception, parallel simulation, and autonomous decision-making. Firstly, an adaptability perception module is established to dynamically assess the alignment between demand and supply, introducing a data circulation mechanism that integrates various data sources for urban management. Given the perceived data, a parallel simulation module is introduced to reproduce multi-scale urban systems. This module simulates dynamic state changes, offering adaptability estimation/predictions for various evolutionary trajectories. Finally, based on the results of high-precision simulations, an autonomous decision-making module is adopted to offer management decisions. The proposed decision-making module enhances the system's overall performance without sacrificing the benefits of individual components. These modules, powered by symbiotic intelligence, harness the combined strengths of humans and machines to enhance urban system agility and flexibility. Four components of the evolutionary city framework are elaborated in the rest part of this section.

\subsection{Symbiotic Intelligence Mechanism}


Symbiotic intelligence is a mechanism for human-machine collaboration. By enhancing alignment and mutual trust between humans and machines, it aims to boost the collaborative efficiency. The human-machine symbiosis mechanism manifests in three collaborative forms: control, cooperation, and organization. Control, at its core, epitomizes a predominantly unidirectional relationship in which humans dictate the actions of agents. Cooperation involves a more reciprocal interaction, necessitating mutual understanding as both humans and agents achieve goals in a coordinated manner. By contrast, organization represents the pinnacle of this symbiosis in the context of evolutionary cities. In this context, humans and intelligent agents intertwine within a complex, unified system, optimized for urban management tasks, exemplifying a highly sophisticated form of partnership. 

Humans and intelligent agents, employing all three aforementioned relationships, can achieve symbiotic intelligence through alignment techniques. Alignment techniques can assist humans and intelligent agents in reaching consensus on values, objectives, and strategies. Specifically, recent advancements in Large Language Models (LLMs) furnish a viable platform for facilitating such alignment. Tools, such as Langchain or analogous platforms, can translate human instructions into machine-executable commands, while concurrently elucidating the system's operational mechanisms to humans. Through such interactions and information exchanges, both humans and intelligent agents can iteratively adapt their behaviors, fostering symbiotic evolution.

To further bolster trust between humans and machines, mitigate unnecessary management delays, and achieve active perception., a multi-level human-machine symbiosis mechanism is employed in the evolutionary city framework. Through the multi-level mechinism, a proactive detection that enabls a comprehensive grasp of urban system dynamics, a parallel simulation that accurately reproduce the system dynamic, and an autonomous decision-making that aligns the values of various decision-makers, can be achieved \cite{lin2023city, wangDAOMetaControlMetaSystems2022}.

\subsection{Adaptability perception}

Adaptability perception denotes how the urban system employs a range of devices to monitor the city's dynamic state in real time. It aims to leverage the fusion and circulation of data to assess the alignment between urban supply and demand.

\begin{figure}[ht]
\centering
\includegraphics[width=0.8\linewidth]{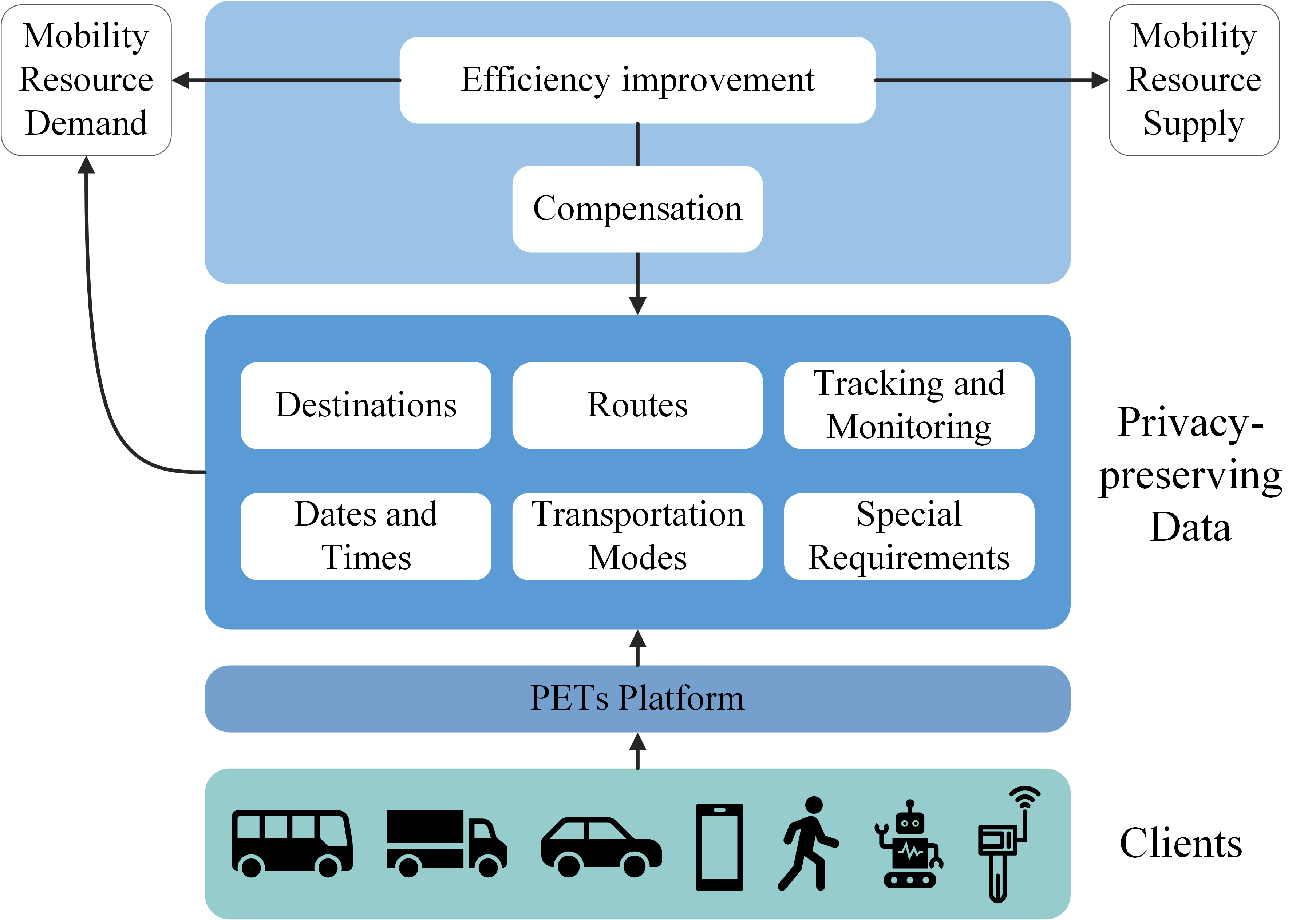}
\caption{City data exchange platform with privacy protection mechanism in adaptability perception.}
\label{sensing}
\end{figure}

In order to support adaptability perception, three key technologies need to be fully applied: multimodal and multi-sources sensors, data privacy protection, and data fusion technology. Sophisticated sensor technology encompasses both sensors set up by city administrators and various devices equipped with sensors owned by city residents. Functioning metaphorically as the city's "nerve endings" \cite{hu2023ir}, these sensors continuously monitor real-time events and conditions within the city. In order to take full advantage of the data collected by sensors, it is essential to provide an efficient data exchange platform while data privacy protection. Only with proper safeguarding of privacy will residents be inclined to share their information with the city's administrative entities. This is where Privacy-Enhancing Technologies (PET) come in, as shown in Fig. \ref{sensing}. With PET, service companies can adjust their offerings based on encrypted data, without seeing private details of individual users. Finally, data fusion techniques are used to integrate the data from different sources through the PET platform. Through data fusion, we can make full use of all aspects of information. These information can complement and confirm each other, and more comprehensively perceive the city's adaptability

\subsection{Parallel simulation}

\begin{figure}[ht]
\centering
\includegraphics[width=0.8\linewidth]{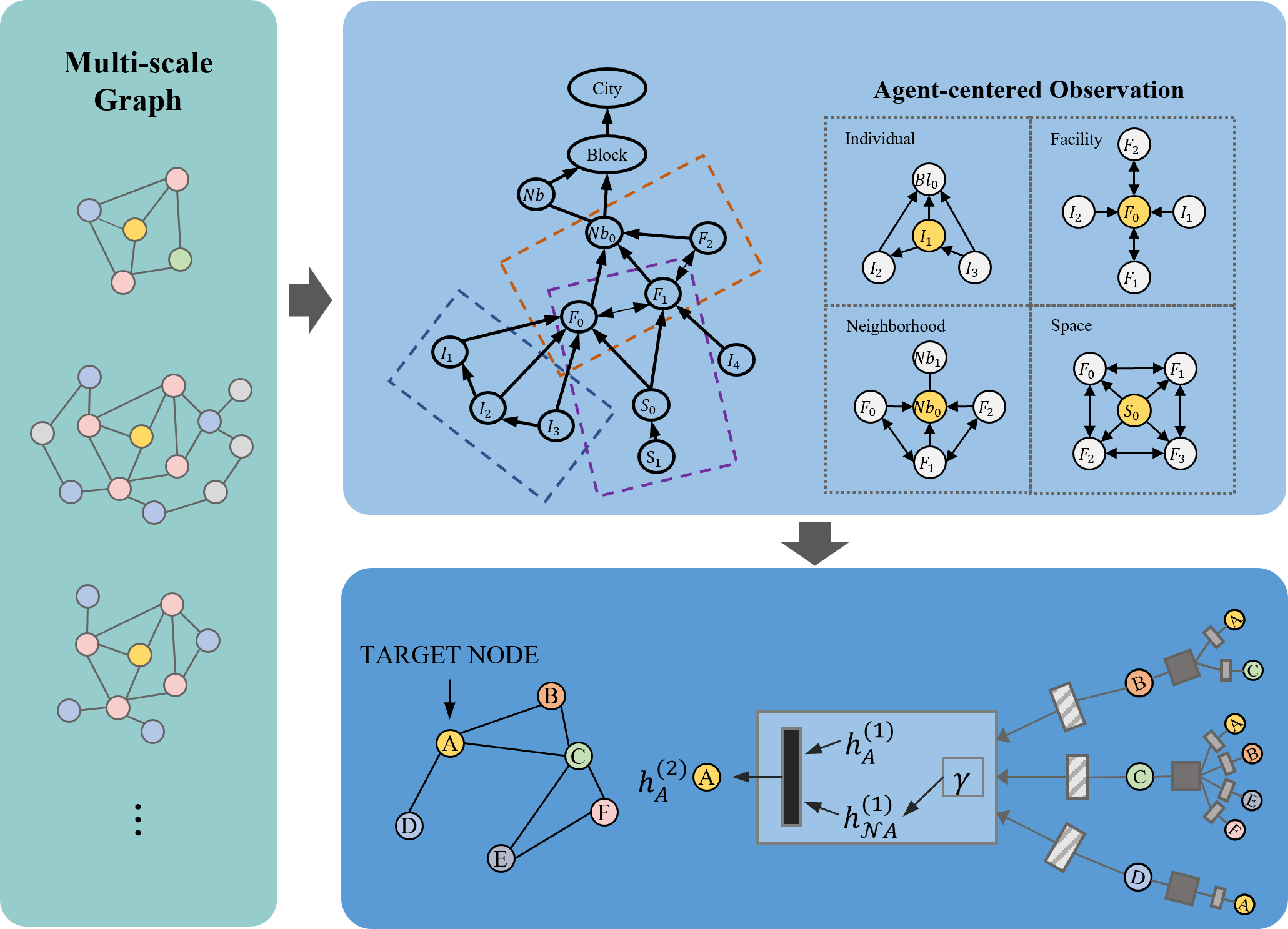}
\caption{City dynamic parallel simulation system based on graph.}
\label{simulation}
\end{figure}

Parallel simulation refers to employing multiple methods to forecast changes in urban adaptability stemming from various evolutionary directions within a virtual space. For this purpose, parallel simulation offers deep insight into urban evolutionary processes by producing varied evolutionary trajectories.

Besides traditional rule-based approaches, one can adopt the emerging technology of data-driven simulation methods \cite{wang2023transworldng}. This simulation method is based on graph neural network (GNN), which encodes components within a city using a graph structure. Additionally, dynamic graph generation algorithms can simulate agents and their interactions within the urban system. The graph-based simulation addresses challenges from heterogeneous agents and their dynamic interactions, offering a versatile solution for modeling intricate urban systems. The architecture of the simulation model is depicted in Fig. \ref{simulation}, with its graph-based and data-driven methodologies being emphasized. Continual input of real-world data at diverse scales enables parallel cross-validation between simulated and real system dynamics. The system's adaptability makes it suitable for a wide range of city resource allocation tasks, positioning it as a powerful tool for crafting realistic city simulations.

\subsection{Autonomous decision-making}

Autonomous decision-making refers to a decision-making mechanism that plentiful intelligent agents apply optimal decisions autonomously without a centralized autority. 

\begin{figure}[ht]
\centering
\includegraphics[width=0.8\linewidth]{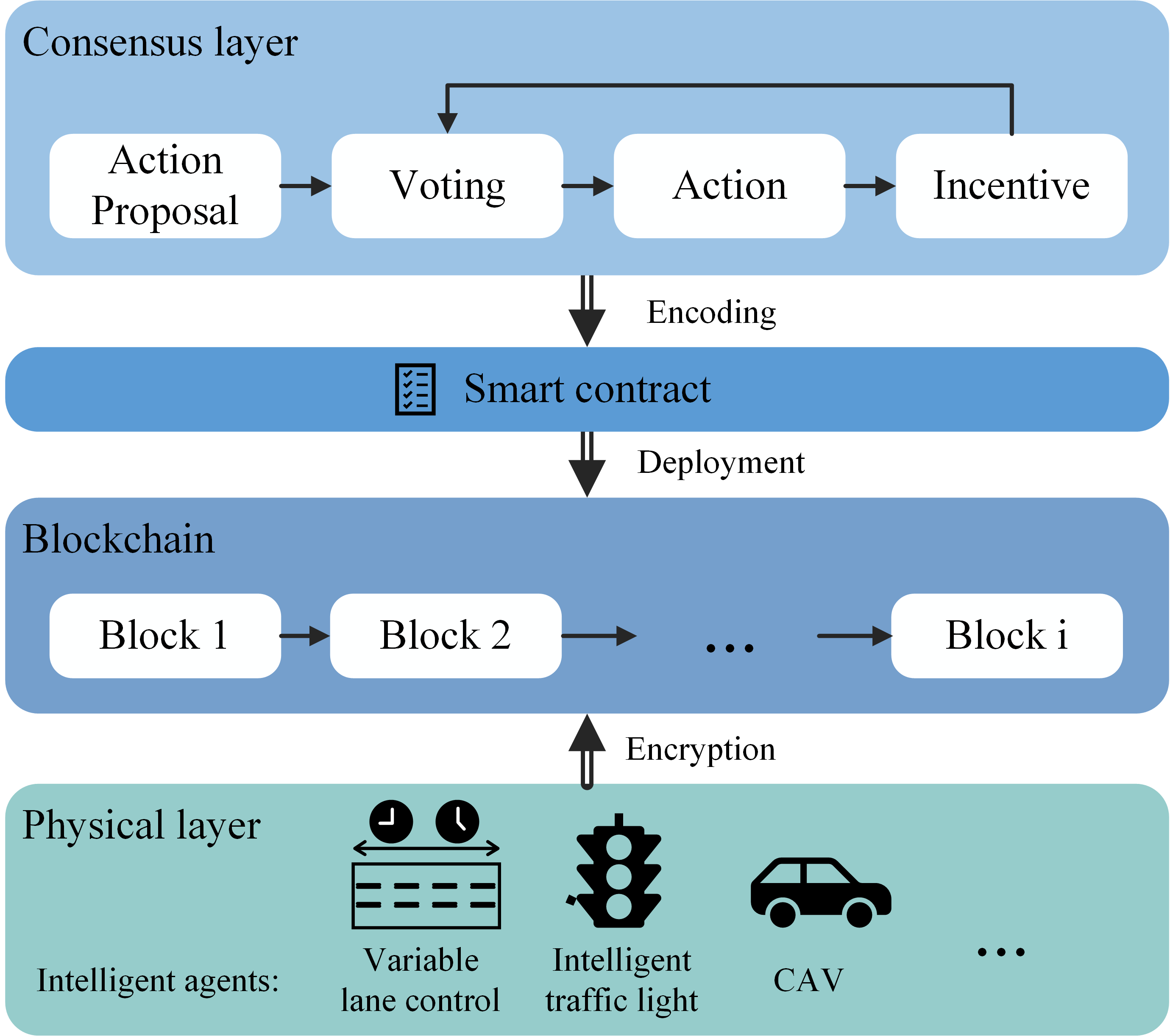}
\caption{Blockchain-based DAO platform for autonomous decision-making.}
\label{DAO}
\end{figure}

The autonomous decision-making mechanism can be achieved through utilizing the  Decentralized Autonomous Organization and Operation (DAOs) system based on the Blockchain technology. As presented by Fig. \ref{DAO}\cite{yao2023towards}, through a proposal-voting-action-incentive procedure, plentiful intelligent agents with divers objectives, granularity and decision frequencies. Without a centralized authority, the DAOs can synchronize the behaviour of intelligent agents by reaching consensus on their individual objectives, meanwhile enhance the overall system performance. Additionally, operating in a decentralized manner, the DAOs improves the usage of computational resources, therefore a effective decision-making process can be achieved.

\section{Case Study} \label{casestudy}

Traffic management is integral to comprehensive urban management. It can span various levels, ranging from micro to macro. Positioned between micro and macro levels, lane management is vital for traffic management, offering strategies to regulate lanes and ensure traffic efficiency. This paper applies the evolutionary city framework to address lane management challenges.

\begin{figure}[ht]
    \centering
    \begin{subfigure}[b]{0.7\linewidth}
        \centering
        \includegraphics[width=1\linewidth]{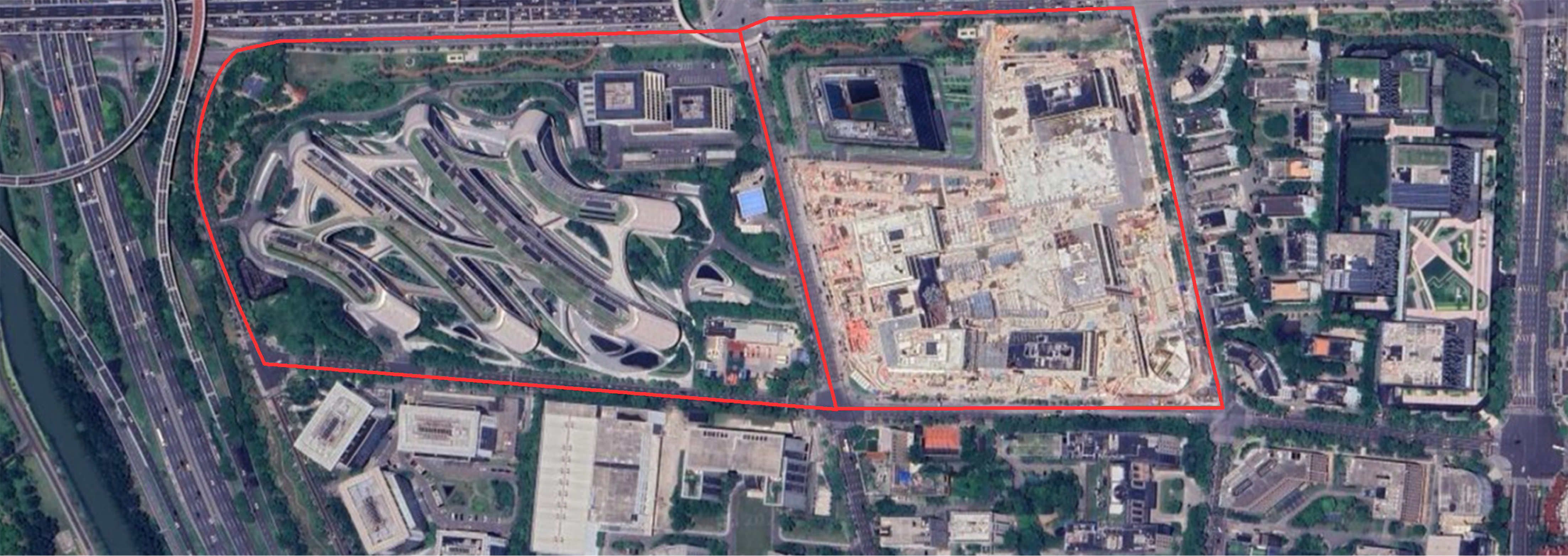}
        \caption{Satellite remote sensing image.}
        \label{remote}
    \end{subfigure}
    \begin{subfigure}[b]{0.7\linewidth}
        \centering
        \includegraphics[width=1\linewidth]{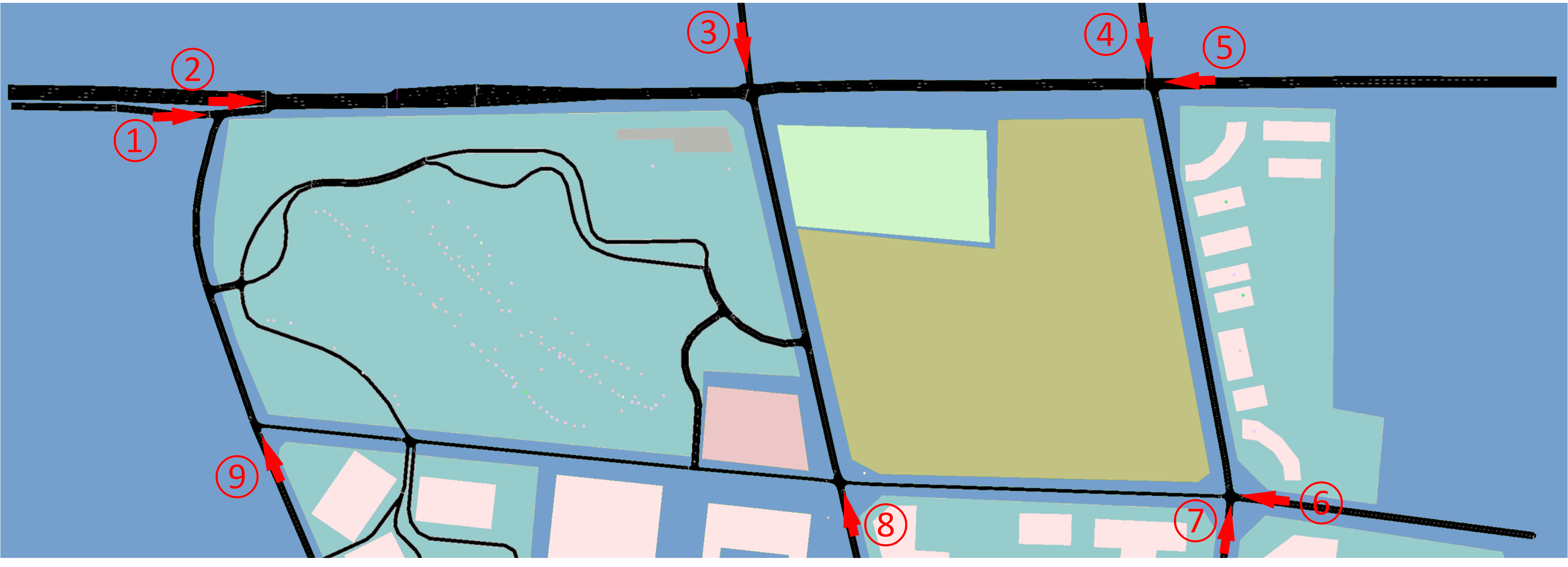}
        \caption{Simulation modeling scenario.}
        \label{vissim}
    \end{subfigure}    
    \caption{Location of the case study. The red line indicates the road where lane control is applied in this experiment}
    \label{sumo}
\end{figure}

\begin{table*}[htbp]
  \centering
  \caption{Morning and evening peak traffic of different access roads in the region.}
  \setlength{\tabcolsep}{3mm}{
    \begin{tabular}{cccccccccc}
    \toprule
    Access Road Number & 1
 & 2 & 3 & 4& 5 & 6 & 7 & 8 & 9 \\
    \midrule
    Traffic Flow in Morning Peak (vehicles /hour) & 1900
 & 440 & 440 & 0 & 550 & 0 & 690 & 550 & 600 \\
    Traffic Flow in Evening Peak (vehicles /hour) & 850
 & 400 & 1000 & 190 & 190 & 60 & 60 & 650 & 850\\
    \bottomrule
    \end{tabular}}
  \label{trafficflow}
\end{table*}


The experimental subject is located in the Shanghai Linkong SOHO Economic Park, as shown in Fig. \ref{vissim}, which is significantly influenced by the complex urban system structure and variations in residents' commuting demands. The traffic flow in the area shows a clear tidal pattern. the different access roads in the area, as noted in numbers 1 to 9, have significant traffic flow differences at different times of the morning and evening peak, as shown in TABLE \ref{trafficflow}. However, the long update cycles of the city's infrastructure hinder its ability to promptly match higher-frequency variations in demand. Therefore, implementing efficient organizational management of lanes is crucial. This strategy becomes an integral part of an evolutionary city, serving as an effective solution to alleviate the mismatch between supply and demand.

\subsection{Evolutionary Lane Management System}

\begin{figure}[ht]
\centering
\includegraphics[width=1\linewidth]{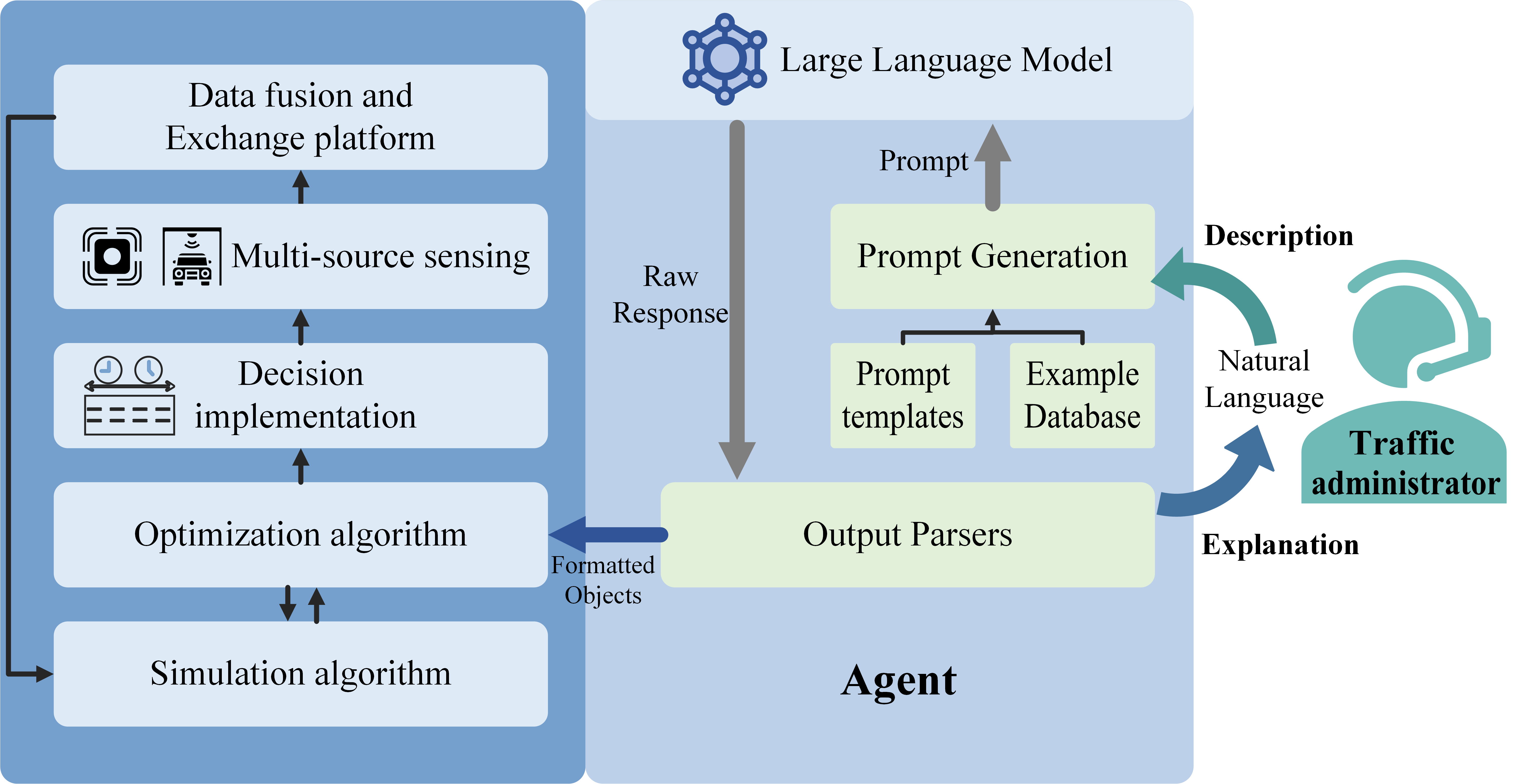}
\caption{Symbiosis intelligent based evolutionary lane management system}
\label{system}
\end{figure}

The system applied in this case operates in a cycle of perception, simulation, and decision-making. The symbiotic intelligence mechanism is implemented for setting lane management objectives, as illustrated in Fig. \ref{system}. Continuous data collection is essential for real-time monitoring of the traffic patterns in the case area. Data is sourced from sensor equipment within this area and feedback from traffic participants. This data offers detailed traffic flow insights for effective management. The data is sent to the traffic management department via a data exchange platform and then analyzed using data fusion techniques. Traffic administrators oversee the real-time traffic state, considering additional factors such as emergency demand and special events, etc. Traffic administrators converse with agents using natural language, aided by large language models. Traffic management objectives are established through iterative, interpretable dialogues. Subsequently, the simulation and decision-making modules refine the management strategy based on these objectives. Once a strategy aligns with the objectives, it's integrated into the intelligent traffic infrastructure. Post-implementation, multi-source sensors persistently monitor the traffic status.




To promote alignment and enhance mutual trust while setting optimization objectives, we facilitate multiple dialogues between the traffic administrator and the agent. We employ a hybrid methodology that integrates the Analytic Hierarchy Process-Entropy Weight Method (AHP-EWM) with LLM. Based on real-time scenarios and additional constraints, the traffic administrator provides a description in natural language. Subsequently, the LLM produces matrices emphasizing the importance of various objectives. Utilizing the generated matrices, we apply the AHP-EWM method to ascertain the weight of each indicator, with each symbolizing a unique optimization objective. We define two primary indicators: road network traffic efficiency ($z_f$) and environmental emissions ($z_n$), with weights $\gamma_f$ and $\gamma_n$, respectively.  For traffic efficiency, we adopt sub-level indicators such as network parking delay $D_s$, average parking frequency per vehicle $C_s$, average delay per vehicle $D_a$, and total delay ($D$). These indicators have associated weights denoted by $\gamma_f^{ds}$, $\gamma_f^{cs}$, $\gamma_f^{a}$, and $\gamma_f^{D}$, respectively. For environmental emissions, the adopted sub-level indicators are $CO_2$ emissions ($E_{CO_2}$), $NO_x$ emissions ($E_{NO_x}$), VOC emissions ($E_{VOC}$), and fuel consumption ($E_f$). These indicators have weights assigned as $\gamma_n^{ec}$, $\gamma_n^{en}$, $\gamma_n^{ev}$, and $\gamma_n^{f}$, respectively. The related mathematical expressions are provided in Equations (\ref{eq:min_PI}).

\begin{equation}\label{eq:min_PI} min PI=\gamma_fz_f+\gamma_nz_n\end{equation}
\[z_f=\gamma_f^{ds}D_s+\gamma_f^{cs}C_s+\gamma_f^aD_a+\gamma_f^DD\]
\[z_n=\gamma_n^{ec}E_{CO_2}+\gamma_n^{en}E_{NO_x}+\gamma_n^{ev}E_{VOC}+\gamma_n^fE_f\]


\begin{figure*}[ht]
\centering
\includegraphics[width=0.8\linewidth]{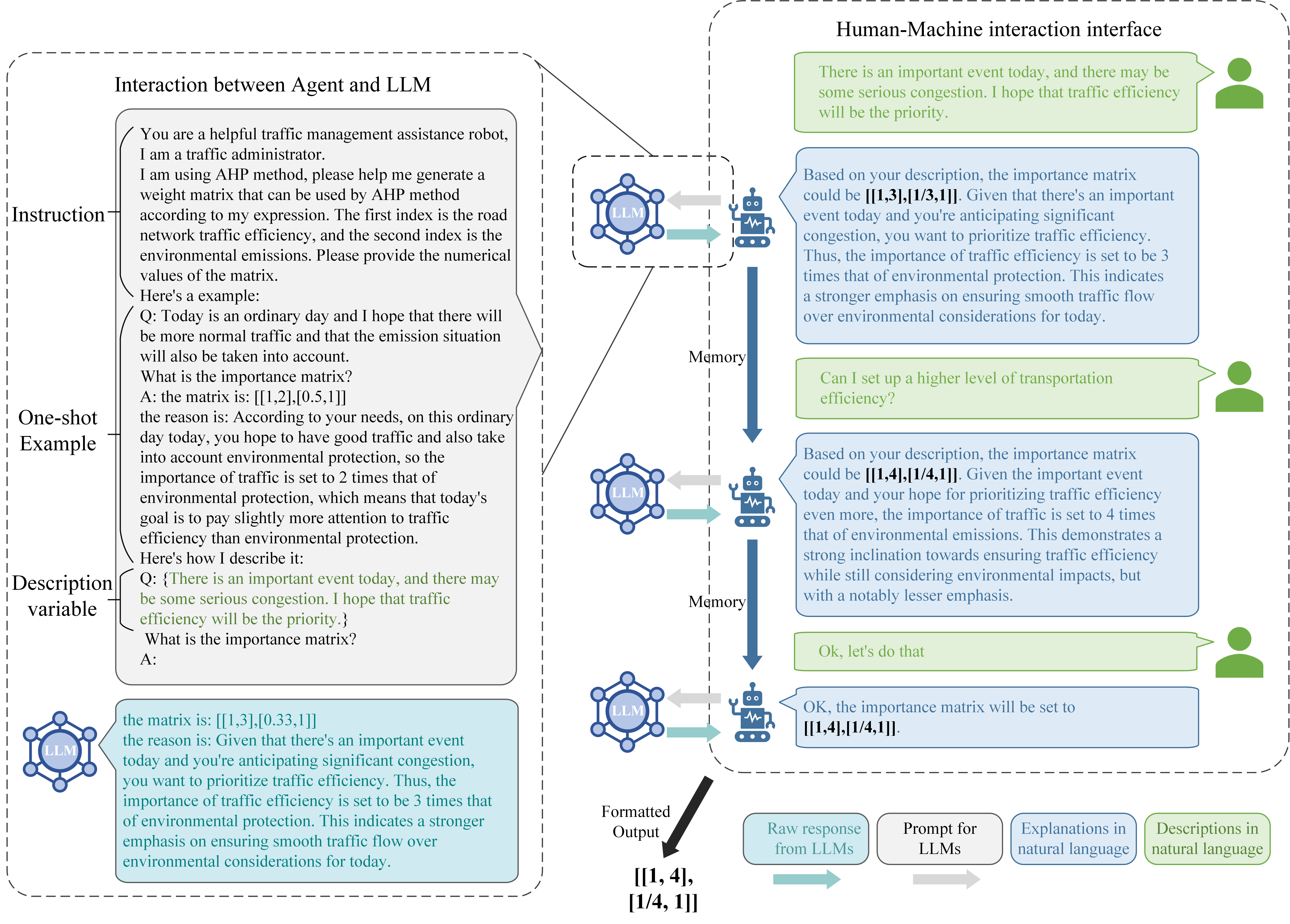}
\caption{The Langchain-driven Agent can chat with the traffic manager in natural language, combined with the prompt template, the LLM can be used to generate the importance matrix available to the machine for subsequent algorithms.}
\label{prompt}
\end{figure*}


To derive an importance matrix from natural language descriptions via LLM, we've developed a Langchain-driven intelligent agent. Based on the traffic administrator's input and a predefined one-shot prompt template, the agent creates an importance matrix and subsequently justifies its reasons to the administrator. The administrator can further engage with the agent to refine the matrix. Once the matrix is finalized to the administrator's satisfaction, the agent forwards it—emphasizing the two primary indicators—to the next algorithm, as depicted in Fig. \ref{prompt}. This procedure can also be adapted for dynamically adjusting the other eight sub-level indicators.

After the subsequent algorithm gets the importance matrix, the matrix was subjected to consistency ratio (CR) tests to calculate the weights for $\gamma_f$ and $\gamma_n$ as shown in Equation (\ref{eq:gamma}). For example, the calculation resulted in a weight ratio of 0.8:0.2 for $\gamma_f$ to $\gamma_n$ using the importance matrix $\begin{pmatrix} 1 & 4 \\ 0.25 & 1 \end{pmatrix}$ generated by LLM as shown in Fig. \ref{prompt}. After the weights of the primary indicators were obtained, the weights of the sub-level indicators were normalized and standardized. The information entropy ($e_{fj}$) for indicators such as network parking delay, average parking frequency per vehicle, average delay per vehicle, and total delay, as well as the information entropy ($e_{nj}$) for indicators such as $CO_2$ emissions, $NO_x$ emissions, VOC emissions, and fuel consumption, were calculated using Equation (\ref{eq:eij}). The information efficiency value ($d_{ij}$) was also computed. These calculations yielded the entropy weights for the sub-level indicators. Furthermore, these dynamically determined quantized weights could then be applied to subsequent simulation and decision algorithms.

\begin{equation}\label{eq:gamma}\gamma_i=\frac{\sum_{j=1}^{n}a_{ij}^{\frac{1}{n}}}{\sum_{k=1}^{n}\prod_{j=1}^{n}a_{kj}^{\frac{1}{n}}}(i=[f,n])\end{equation}
\begin{equation}\label{eq:eij}e_{ij}=-\frac{1}{lnn}\sum_{i=1}^{n}p_{ij}ln{(}p_{ij})d_{ij}(i=[f,n],j=1,2,...,m)\end{equation}

In addition to lane control, lane signal control optimization is also necessary. The optimization of signal timing for multiple intersections adopts the objective of minimizing the average vehicle delay time, and an optimization function is established accordingly. To improve the delay estimation in the Webster method, the average delay of each vehicle at the intersection is denoted as $d$ seconds, as shown in Equation (\ref{eq:d}).

\[
    d^\ast=min~d 
\]
\begin{align}
\label{eq:d}
d = \frac{\sum_{i=1}^{n} \sum_{j=1}^{m} \sum_{k=1}^{p} \left[\frac{C(1+\lambda_{ik})}{2(1-\lambda x_{ijk})} + \frac{x_{ijk}^2}{2q_{ijk}^\ast(1-x_{ijk})}\right]\cdot q_{ijk}}{\sum_{i=1}^{n} \sum_{j=1}^{m} \sum_{k=1}^{p} q_{ijk}}
\end{align}
s.t.:
\begin{equation}\label{eq:ti}\sum_{i}{t_i=}C-L\end{equation}
\begin{equation}\label{eq:tii}t_i=Cy_{imax}/x_{max}\end{equation}


Equation (\ref{eq:ti}) represents the constraint for effective green time, where $t_i$ denotes the duration of phase $i$, and $L$ represents the total loss time. Equation (\ref{eq:tii}) represents the constraint for maximum saturation, where $y_{imax}$ is the flow ratio at maximum saturation. 

\begin{figure}[ht]
\centering
\includegraphics[width=0.9\linewidth]{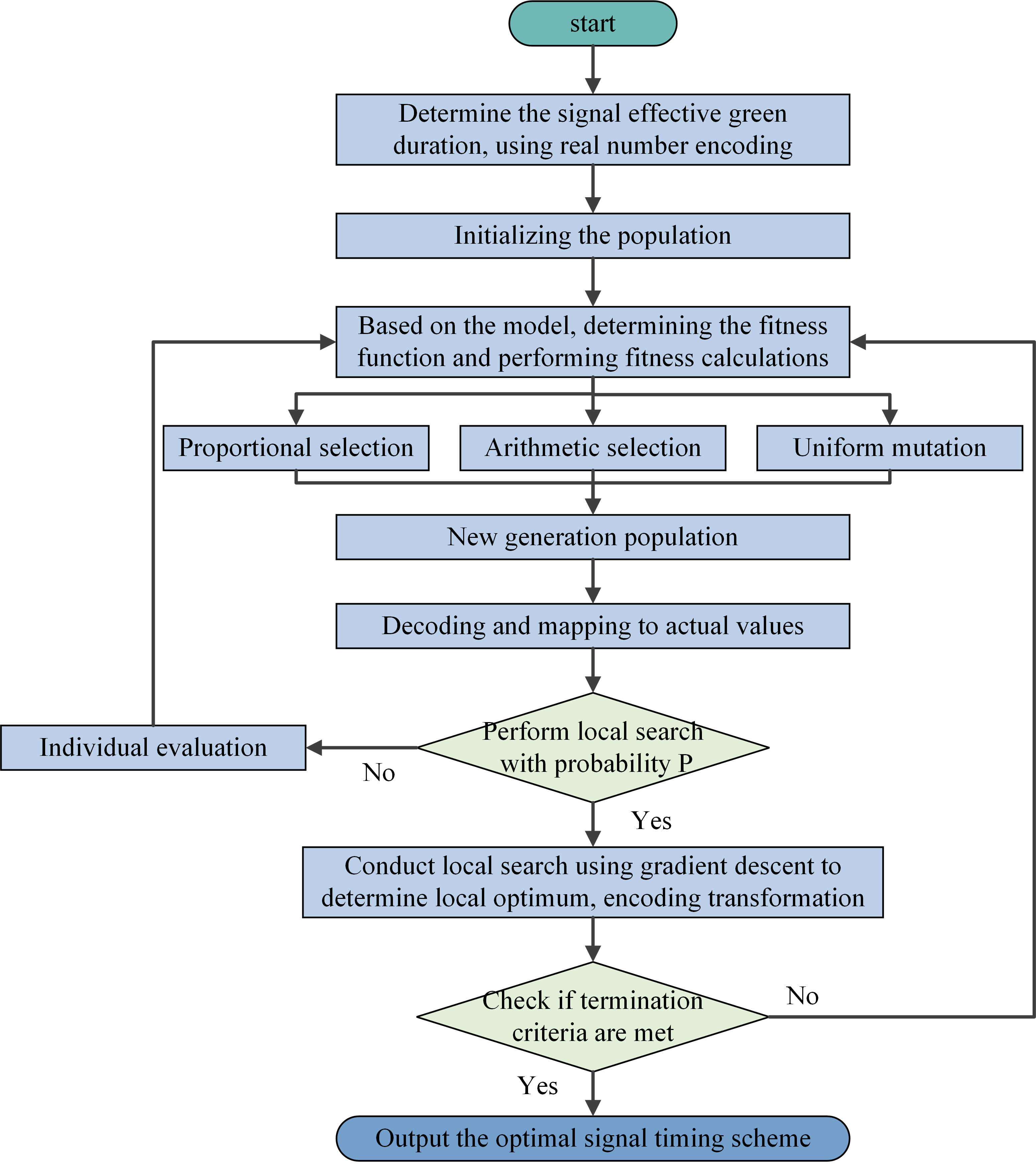}
\caption{Optimization of signal timing flow chart by evolutionary-alike algorithm}
\label{ga}
\end{figure}


An integrated control is applied following the DAO-based control mechanism in Fig. \ref{DAO}. The core of the proposed DAO-based mechanism is the design of the smart contract. This smart contract is used to coordinate and control the voting weights of all agents. To this end, smart contracts need to be simulated and evaluated for evolutionary direction. In the smart contract, the evolutionary-alike evolutionary algorithm is responsible for giving the evolutionary direction of the system based on the optimal decision, and the evolutionary mechanism of the system is derived by accurate simulation. Specifically, Vissim based on rules and the TransworldNG \cite{wang2023transworldng} based on Graph neural networks were used as high-precision simulations. On the basis of high-precision dynamic simulation, the evolutionary algorithm is used to optimize the signal control plan. This algorithm adopts real-value encoding and defines its own fitness function. Through the use of selection, crossover, and mutation operations, the algorithm boosts computational efficiency and identifies the best solution. The detailed computational workflow is shown in Fig. \ref{ga}.

All intelligent agents (i.e., traffic signals in this case study) perform on the basis of their own control logic regarding their particular system dynamics. By encrypting these agents on the Blockchain, all proposed control efforts will be compared with the optimized results from the deployed smart contract. Afterward, the voting power of each agent will be adjusted and the applied control effort of each agent will be determined through the voting and consensus mechanism.

\subsection{Result}
\subsubsection{Optimization results of variable lane cluster control}

The evaluation results of the optimized control for the variable lane area in an arterial road network are presented in TABLE \ref{tab:table7}. Comparing the performance of the area with variable lanes to the original condition of the arterial road, both the network traffic efficiency and environmental emissions show a decrease. However, after implementing group control of variable lanes and optimizing signal timing, the performance further improved significantly. In particular, the improvement during the morning peak period is greater than that during the evening peak period, with a reduction of 17.46\% in network traffic efficiency and 20.53\% in environmental emissions. These results demonstrate a clear improvement in the effectiveness of the measures.

\begin{table*}[htbp]
  \centering
  \caption{Result of evaluation on Linkong variable lane network}
  \setlength{\tabcolsep}{3mm}{
    \begin{tabular}{ccccc}
    \toprule
    Road section name & evaluation item
 & original state$\to$current state & current state$\to$postoptimality &  \\
    \midrule
    Linkong variable lane network & Traffic efficiency & 8.83\% & 7.27\% & 17.46\% \\
          & environmental emission & 4.66\% & 9.28\% & 20.53\% \\
    \bottomrule
    \end{tabular}}
  \label{tab:table7}
\end{table*}

\subsubsection{Comparison of single road optimization and multi-variable lane cluster optimization results}



\begin{table*}[htbp]
\centering
\caption{Comparison Scenario Code Reference Table}
\label{tab:compare-index}
\setlength{\tabcolsep}{5mm}{
\begin{tabular}{|c|cc|cc|}
\hline
Code                    & \multicolumn{1}{c|}{1}   & 2   & \multicolumn{1}{c|}{3}   & 4   \\ \hline
\multirow{2}{*}{Comparison Scenario} & \multicolumn{2}{c|}{original state→current state}     & \multicolumn{2}{c|}{current state→postoptimality
}    \\ \cline{2-5} 
                      & \multicolumn{1}{c|}{morning peak} & evening peak
 & \multicolumn{1}{l|}{morning peak} & evening peak
 \\ \cline{1-5}
\end{tabular}}
\end{table*}

\begin{figure}[ht]
    \centering
    \begin{subfigure}[b]{0.9\linewidth}
        \centering
        \includegraphics[width=1\linewidth]{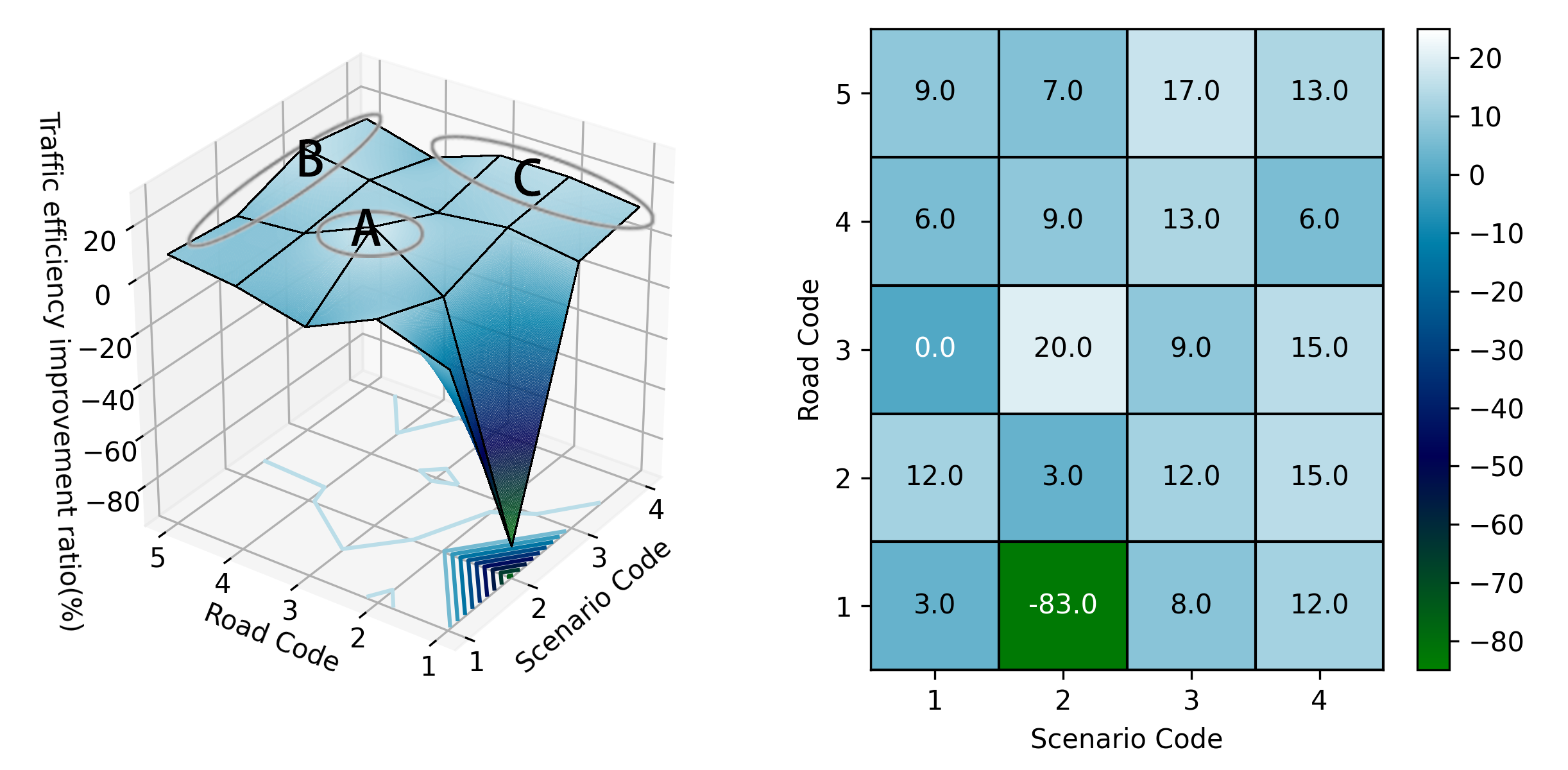}
        \caption{Traffic efficiency optimization results.}
        \label{r1}
    \end{subfigure}
    \begin{subfigure}[b]{0.9\linewidth}
        \centering
        \includegraphics[width=1\linewidth]{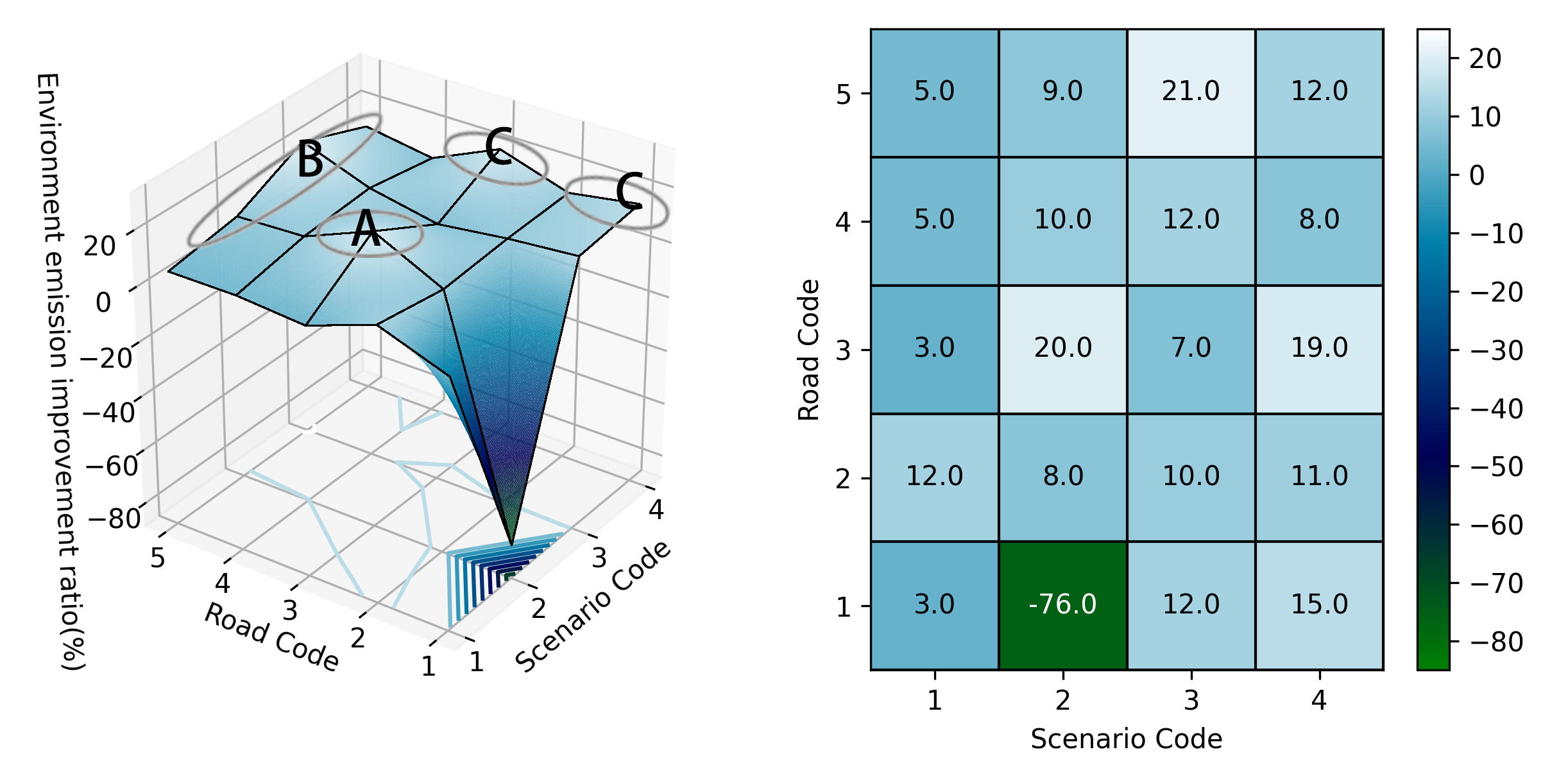}
        \caption{Environmental emission optimization results.}
        \label{r2}
    \end{subfigure}    
    \caption{The result of variable lane optimization. Road Code: 1. Guangshun North Road, 2. Xiehe Road, 3. Fuquan Road, 4. Jinzhong Road, 5. Linkong variable lane network. Scenario Code shown in TABLE \ref{tab:compare-index}.}
    \label{r}
\end{figure}

The optimization results of single road and variable lane group control were compared and analyzed. The comparison results of traffic efficiency and environmental emission status are shown in Figure \ref{r1} and \ref{r2} respectively.  Observations from A, B, and C reveal a notable improvement, up to 20\%, in the traffic efficiency and environmental emission status on Fuquan Road, Jinzhong Road, and Lingkong Road. It should be pointed out that in the evening rush hour, the traffic efficiency of the section of Guangshun Road decreased and the emission status deteriorated, mainly because the signal control was increased at the intersection after the original one-way street was changed to the two-way variable lane. However, the change to the two-way variable lane significantly reduced the detour of vehicles in the region and reduced the overall environmental pollution in the region.

\begin{figure}[ht]
    \centering
    \begin{subfigure}[b]{0.8\linewidth}
        \centering
        \includegraphics[width=1\linewidth]{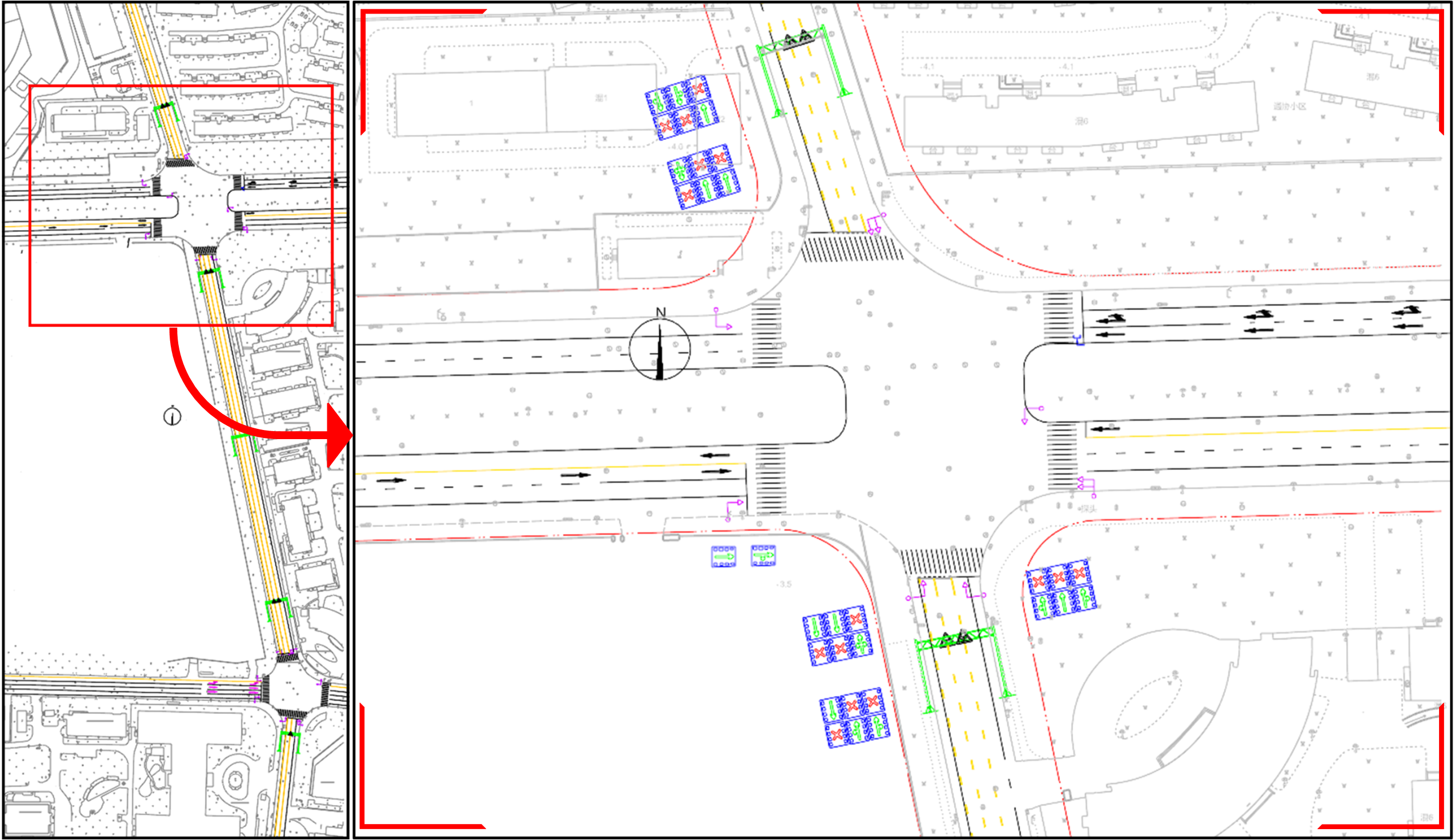}
        \caption{Design drawing of the variable lane facility at Fuquan Road intersection, the green block shows the gantry for managing the variable lane, and the blue block shows the control signal plans.}
        \label{cad1}
    \end{subfigure}
    \begin{subfigure}[b]{0.8\linewidth}
        \centering
        \includegraphics[width=1\linewidth]{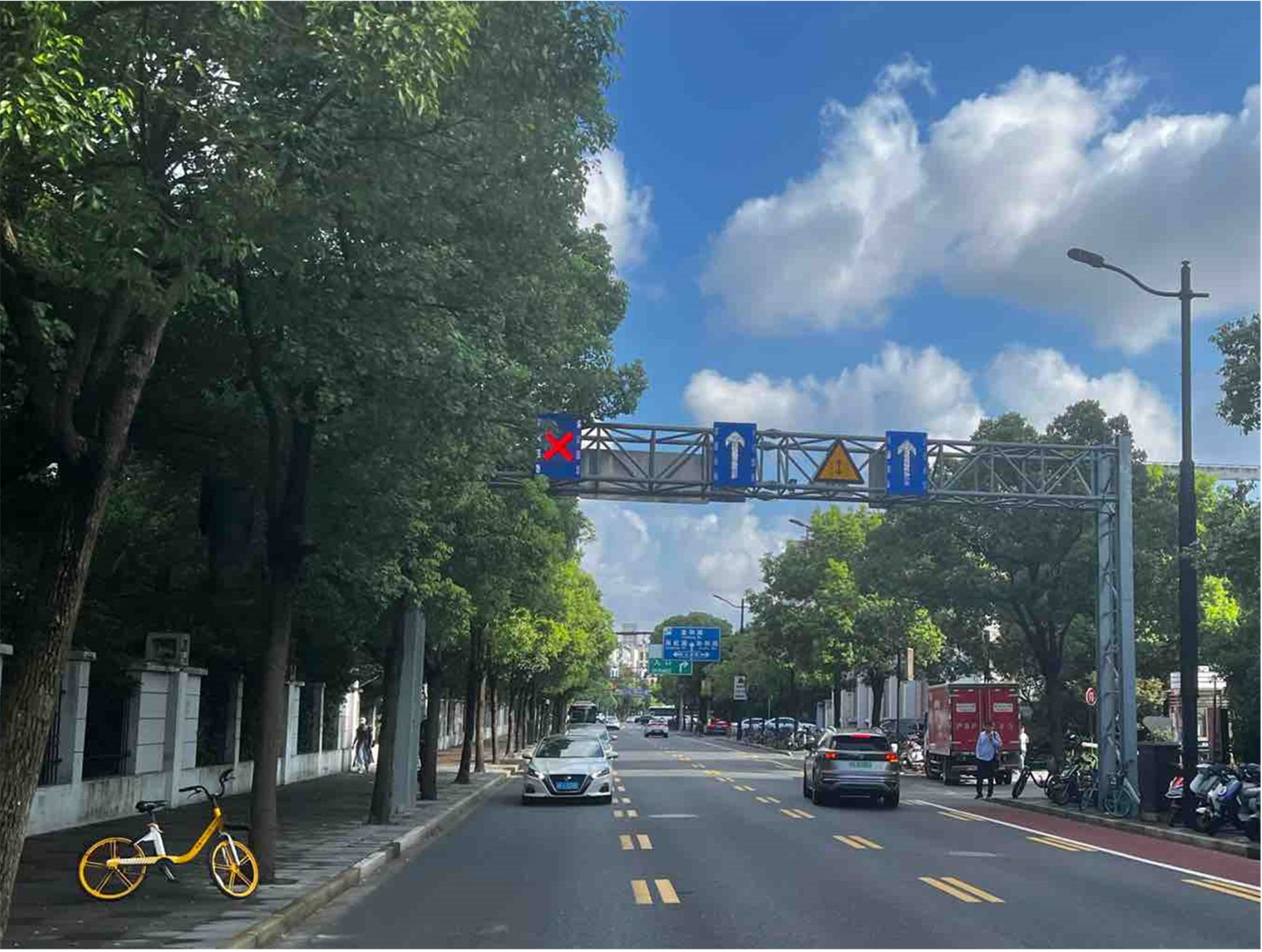}
        \caption{Photo of the variable lane management facility after completion.}
        \label{cad2}
    \end{subfigure}    
    \caption{Fuquan Road intersection design and photos after construction.}
    \label{cad}
\end{figure}

Finally, we put our variable lane management scheme into practice. As shown in Fig. \ref{cad} is the design drawing of Fuquan Road intersection and the photo taken after the completion of construction. According to the statistical results after implementation, it is shown that our method can reduce the travel delay of this road by about 15\%-20\%.
\section{Conclusion and future works} \label{conclusion}

In conclusion, this paper proposes a conceptual framework of an evolutionary city along with a methodological approach to enhance the agility and flexibility of urban systems in response to constantly changing demands. 
This approach enables prediction and scheduling based on the dynamic demands of cities, leading to more flexible and agile city development.  
Simulation experiments and application of the framework to city lane management have demonstrated 
a significant reduction in traffic delays by 15\%-20\%.
By incorporating symbiotic intelligence and considering human factors, 
it becomes possible to effectively perceive the dynamic demand, simulate urban evolution, and make informed decisions that align with the values and preferences of their communities in urban management and planning.

However, there are still limitations that need to be addressed, such as the need for extensive case validation and the exploration of collaborative mechanisms between artificial intelligence and human experts. Future research will focus on enhancing the accuracy of simulation and prediction, validating the effectiveness of the framework in a more comprehensive manner, and validating the symbiotic intelligence mechanism in a large-scale scenario. Additionally, further exploration is warranted in areas such as the concept of "free-lane city" \cite{malekzadeh2023overlapping-freelane1,malekzadeh2021linear-freelane2}, which aligns more closely with future urban scenarios and embraces the principles of flexibility and symbiosis. Inspired by Transportation 5.0\cite{8317905}, future urban planning and design concepts should be tailored to the functional and value-based needs of future cities, adopting more flexible principles. 

Overall, the concept of evolutionary cities provides new ideas and technological means for the autonomous evolution of cities. This research is of great significance in improving the sustainability and resilience of the urban system, achieving harmonious development between humans and the environment, and ultimately achieving a symbiotic evolution of the urban systems. 



\ifCLASSOPTIONcaptionsoff
  \newpage
\fi




\bibliographystyle{IEEEtran}
\bibliography{./IEEEexample}

\vspace{-1cm}

\begin{IEEEbiography}
[{\includegraphics[width=1in,height=1.25in,clip,keepaspectratio]{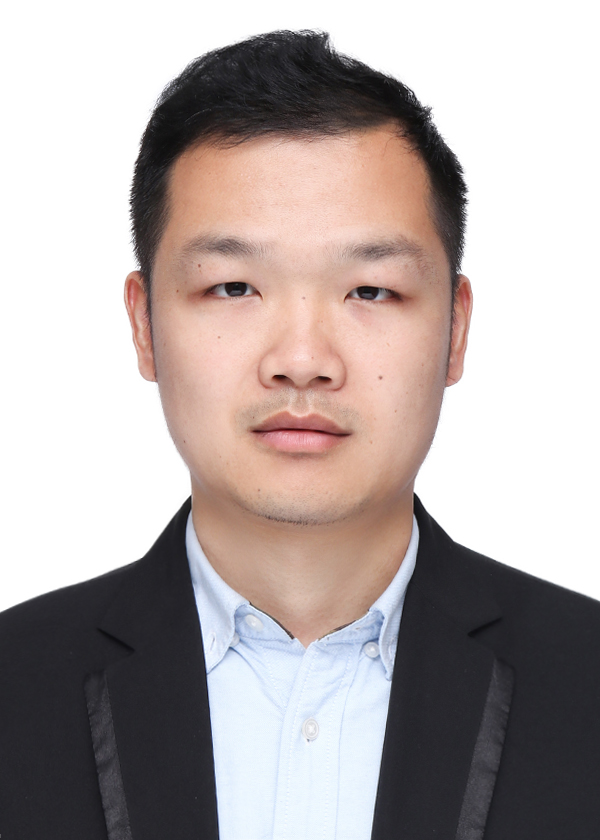}}]
{Xi Chen} works as advisor at Shanghai Artificial Intelligence Laboratory and Deputy Chief Engineer at Shanghai Urban Construction Design \& Research Institute (Group) Co., Ltd., known as Huanjiao Institute. He has been honored with the Excellent Paper Award at the 14th World Congress on Intelligent Transport Systems and the Second Prize for Scientific and Technological Progress by the China Highway and Transportation Society. He has also received numerous national and Shanghai awards for outstanding engineering survey and design. He has been extensively involved in scientific research, consulting, design, and EPC implementation in the fields of intelligent transportation, urban renewal, and smart cities. He has led or participated in over 10 scientific research projects (including 2 national science and technology programs) and more than 100 design projects. He has published over 30 academic papers.
\end{IEEEbiography}

\vspace{-1cm}
\begin{IEEEbiography}
[{\includegraphics[width=1in,height=1.25in,clip,keepaspectratio]{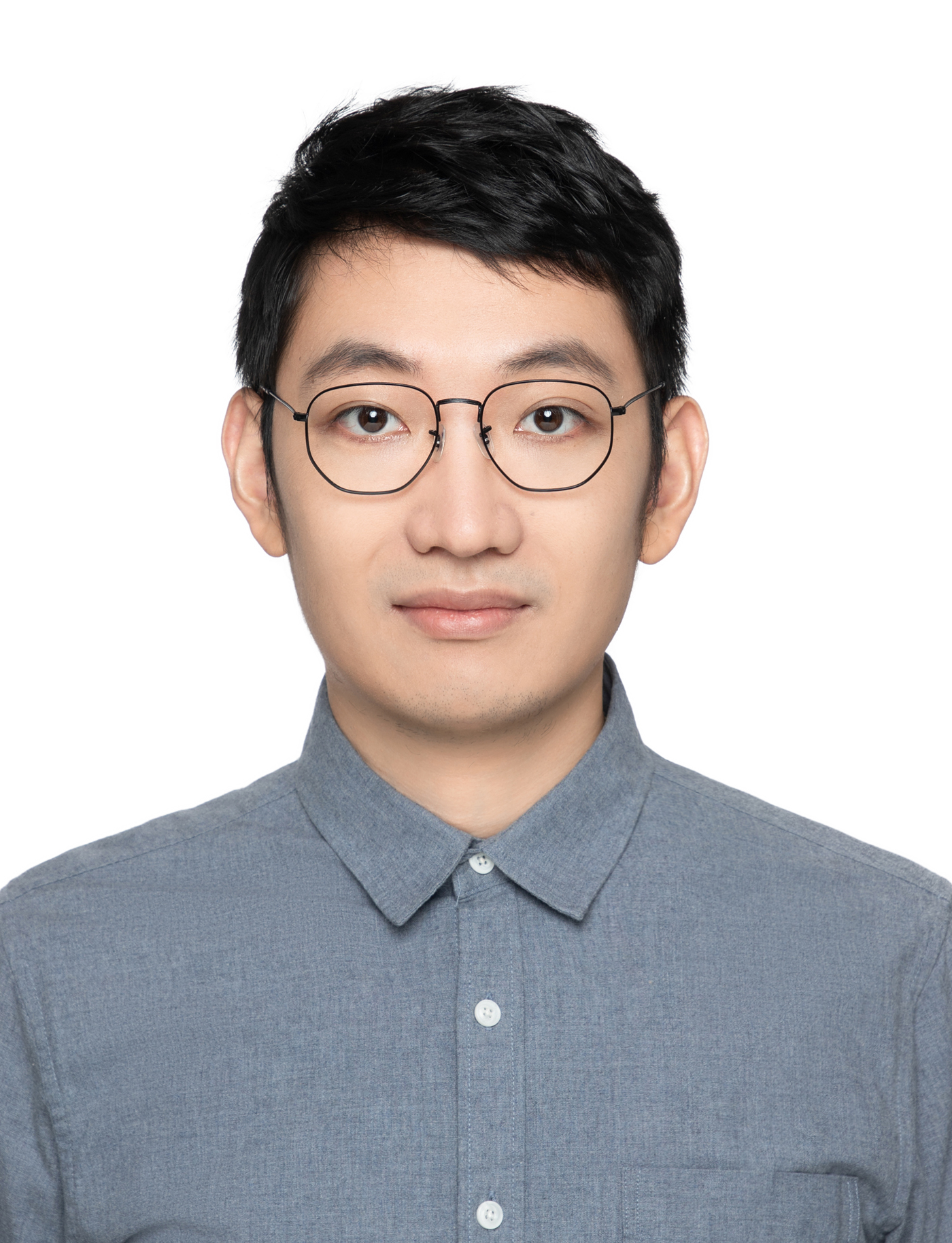}}]
{Wei Hu} received his Ph.D. degree in Architecture from College of Architecture and Urban Planning of Tongji University in 2023, Shanghai, China. After graduation, he joined the Shanghai Artificial Intelligence Laboratory in Shanghai, China as a researcher. His current research focuses are centered on artificial intelligence, computer aided architecture design (CAAD),social simulation, and intelligent transportation system. He served as an individual reviewer of  IEEE transactions on IV and multiple CAAD conferences including ITSC and CAADRIA.
\end{IEEEbiography}
\vspace{-1cm}
\begin{IEEEbiography}
[{\includegraphics[width=1in,height=1.25in,clip,keepaspectratio]{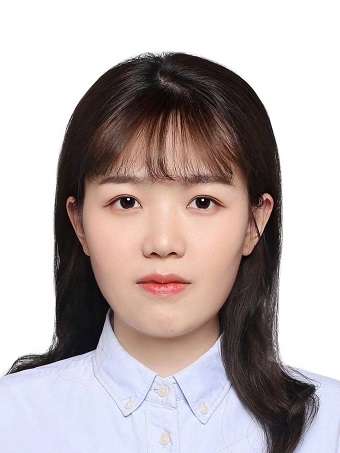}}]
{Jingru Yu} received her Ph.D. degree in Highway and Transportation Engineering from Zhejiang University, China, in 2022.  After graduation, she joined the Shanghai Artificial Intelligence Laboratory in Shanghai, China as a research associate. Her current research focuses are centered on artificial intelligence, autonomous driving, and intelligent transportation system. She served as an individual reviewer of  IEEE transactions on IV and multiple IEEE conferences including ITSC and IEEE SMC.
\end{IEEEbiography}
\vspace{-1cm}
\begin{IEEEbiography}
[{\includegraphics[width=1in,height=1.25in,clip,keepaspectratio]{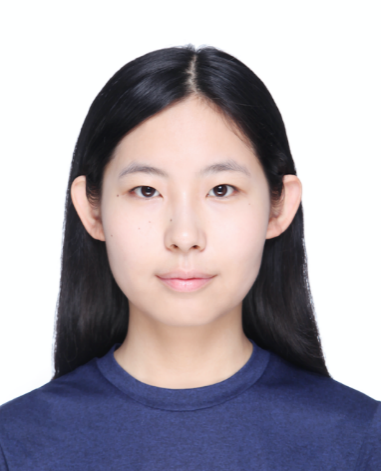}}]
{Ding Wang} received her Ph.D. degree in Transportation Planning and Engineering from New York University (NYU) in 2022, NY, USA. From 2017 to 2022, she served as a research assistant at NYU's C2SMART University Transportation Center. After graduation, she joined the Shanghai Artificial Intelligence Laboratory in Shanghai, China, with the support of the Shanghai Young Talent Program. She has been a key contributor to the development of the urban traffic virtual simulation platform, MATSim-NYC. Her current research focuses are centered on artificial intelligence, data-driven traffic modeling and simulation, traffic emergency management, and intelligent transportation system.
\end{IEEEbiography}
\vspace{-1cm}

\begin{IEEEbiography}
[{\includegraphics[width=1in,height=1.25in,clip,keepaspectratio]{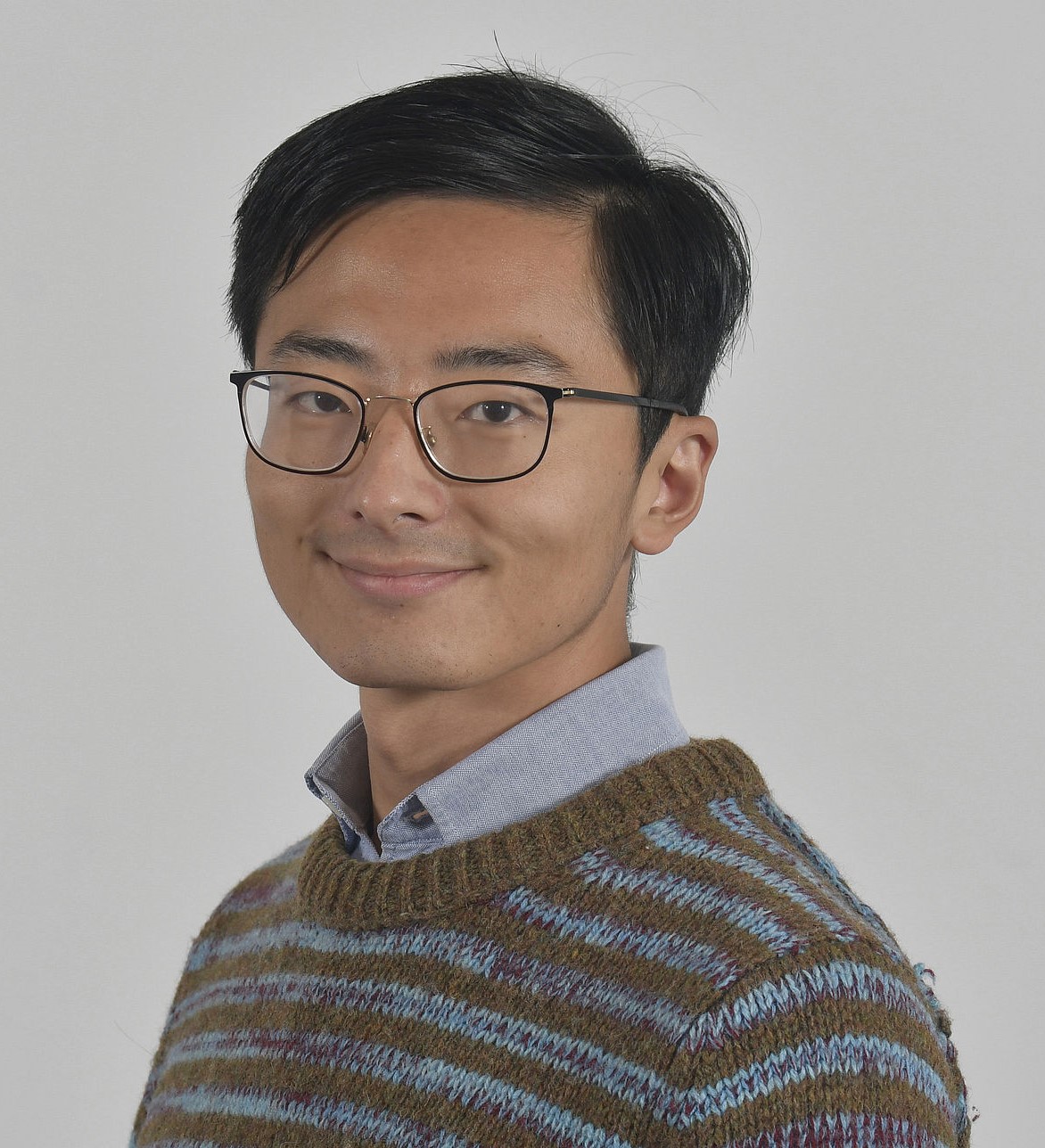}}]
{Shengyue Yao} is currently a researcher at Shanghai Artificial Intelligence Laboratory. He is working towards his Ph.D.degree in Transportation and Urban Engineering, Technical University of Braunschweig, Germany. His research interests include integrated control on complex system, intelligent transportation system, and urban resource management.He served as an individual reviewer of IEEE transaction on ITS, IEEE transaction on IV and multiple IEEE conferences including ITSC and IEEE SMC.
\end{IEEEbiography}
\vspace{-1cm}

\begin{IEEEbiography}
[{\includegraphics[width=1in,height=1.25in,clip,keepaspectratio]{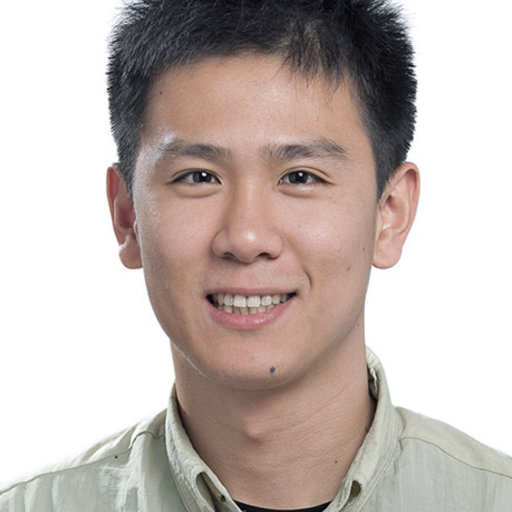}}]
{Yilun Lin} (Member, IEEE) received his Ph.D.degree in control science and engineering from The State Key Laboratory of Management and Control for Complex Systems, Institute of Automation, Chinese Academy of Sciences (CASIA), China, in 2019. 
He worked as an Assistant Professor at the State Key Laboratory for Management and Control of Complex Systems, Chinese Academy of Sciences, from 2019 to 2020, and as a Senior Algorithm Engineer at Ant Group from 2020 to 2022. 
He is currently a Research Scientist and the Principal Investigator of Urban Computing Lab, Shanghai AI Laboratory. His research interests include social computing, urban computing, intelligent transportation systems, deep learning, reinforcement learning and privacy-preserving computing.
\end{IEEEbiography}
\vspace{-1cm}

\begin{IEEEbiography}
[{\includegraphics[width=1in,height=1.25in,clip,keepaspectratio]{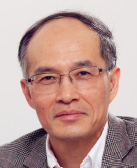}}]
{Fei-Yue Wang} (Fellow, IEEE) received his Ph.D. degree in computer and systems engineering from the Rensselaer Polytechnic Institute, Troy, NY, USA, in 1990. He joined The University of Arizona in 1990 and became a Professor and the Director of the Robotics and Automation Laboratory and the Program in Advanced Research for Complex Systems. In 1999, he founded the Intelligent Control and Systems Engineering Center at the Institute of Automation, Chinese Academy of Sciences (CAS), Beijing, China, under the support of the Outstanding
Chinese Talents Program from the State Planning Council, and in 2002, was appointed as the Director of the Key Laboratory for Complex Systems and Intelligence Science, CAS, and Vice President of Institute of Automation, CAS in 2006. He found CAS Center for Social Computing and Parallel Management in 2008, and became the State Specially Appointed Expert and the Founding Director of the State Key Laboratory for Management and Control of Complex Systems in 2011. His current research focuses on methods and applications for parallel intelligence, social computing, and knowledge automation. He is a Fellow of INCOSE, IFAC, ASME, and AAAS. In 2007, he received the National Prize in Natural Sciences of China, numerous best papers awards from IEEE Transactions, and became an Outstanding Scientist of ACM for his work in intelligent control and social computing. He received the IEEE ITS Outstanding Application and Research Awards in 2009, 2011, and 2015, respectively, the IEEE SMC Norbert Wiener Award in 2014, and became the IFAC Pavel J. Nowacki Distinguished Lecturer in 2021.

Since 1997, he has been serving as the General or Program Chair of over 30 IEEE, INFORMS, IFAC, ACM, and ASME conferences. He was the President of the IEEE ITS Society from 2005 to 2007, the IEEE Council of RFID from 2019 to 2021, the Chinese Association for Science and Technology, USA, in 2005, the American Zhu Kezhen Education Foundation from 2007 to 2008, the Vice President of the ACM China Council from 2010 to 2011, the Vice President and the Secretary General of the Chinese Association of Automation from 2008 to 2018, the Vice President of IEEE Systems, Man, and Cybernetics Society from 2019 to 2021. He was the Founding Editor-in- Chief (EiC) of the International Journal of Intelligent Control and Systems from 1995 to 2000, IEEE ITS Magazine from 2006 to 2007, IEEE/CAA JOURNAL OF AUTOMATICA SINICA from 2014-2017, China’s Journal of Command and Control from 2015-2021, and China’s Journal of Intelligent Science and Technology from 2019 to 2021. He was the EiC of the IEEE Intelligent Systems from 2009 to 2012, IEEE TRANSACTIONS on Intelligent Transportation Systems from 2009 to 2016, IEEE TRANSACTIONS ON COMPUTATIONAL Social Systems from 2017 to 2020. Currently, he is the President of CAA’s Supervision Council, and the EiC of IEEE Transaction on Intelligent Vehicles.
\end{IEEEbiography}


 





\end{document}